\documentclass[showpacs,onecolumn,prd,nofootinbib]{revtex4-1}
 \usepackage{color,graphicx,epsfig}
 \usepackage{amsmath,amssymb,dsfont,epsfig,graphicx,xcolor}
 \usepackage{ifpdf}
 \usepackage{amsmath}
 \usepackage{bm}
 \usepackage{color}
 \usepackage[english]{babel}
 \usepackage{graphicx}%
 \usepackage{amsfonts}%
 \usepackage{amssymb}
 \usepackage{braket}
 \usepackage{hyperref}
 \usepackage{epstopdf}
 \graphicspath{{Plots/}}
\usepackage[utf8]{inputenc}

\newtheorem{notation}{Notation}
\newtheorem{definition}{Definition}
\newtheorem{theorem}{Theorem}

\definecolor{nicered}{rgb}{0.7,0.1,0.1}
\definecolor{nicegreen}{rgb}{0.1,0.5,0.1}
\definecolor{violet}{rgb}{0.7,0.3,0.3}
\hypersetup{colorlinks,citecolor= nicegreen,linkcolor= nicered}

\newcommand{\nc}{\newcommand}
\nc{\non}{\nonumber}
\nc{\hc}{\hbox {H.c.}}
\nc{\noi}{\noindent}
\nc{\barx}{\bar{x}}
\nc{\pbarn}{\;\hbox {pb}}
\nc{\fbarn}{\;\hbox {fb}}

\nc{\hsp}{\hspace{0.5cm}}
\nc{\lsp}{\hspace{1cm}}
\nc{\Lsp}{\hspace{2cm}}
\nc{\LLsp}{\lsp\lsp}
\nc{\lra}{\longrightarrow}
\nc{\p}{\prime}
\nc{\sgn}{\text{sgn}}
\nc{\ph}{\varphi}
\nc{\op}{{\cal O}}
\nc{\eq}{\text{Eq.~}}
\nc{\beq}{\begin{equation}}  \nc{\eeq}{\end{equation}}
\nc{\bea}{\begin{eqnarray}}  \nc{\eea}{\end{eqnarray}}
\nc{\baa}{\begin{array}}     \nc{\eaa}{\end{array}}
\nc{\bit}{\begin{itemize}}   \nc{\eit}{\end{itemize}}
\nc{\ben}{\begin{enumerate}} \nc{\een}{\end{enumerate}}
\nc{\bce}{\begin{center}}    \nc{\ece}{\end{center}}
\nc{\bpm}{\begin{pmatrix}}   \nc{\epm}{\end{pmatrix}}
\nc{\bvt}{\begin{verbatim}}  \nc{\evt}{\end{verbatim}}

\begin{document}

\def\LjubljanaFMF{Faculty of Mathematics and Physics, University of Ljubljana,
 Jadranska 19, 1000 Ljubljana, Slovenia }
\def\LjubljanaIJS{Jo\v zef Stefan Institute, Jamova 39, 1000 Ljubljana, Slovenia}

\title{Light Majorana Neutrinos in (Semi)invisible Meson Decays}
 
\author{Bla\v z~Bortolato}
\email[Electronic address:]{blaz.bortolato@ijs.si} 
\affiliation{\LjubljanaIJS}

\author{Jernej~F.~Kamenik}
\email[Electronic address:]{jernej.kamenik@cern.ch} 
\affiliation{\LjubljanaIJS}
\affiliation{\LjubljanaFMF}

\begin{abstract}

We reconsider decays of pseudoscalar mesons ($P$) to neutrino pairs and possibly additional photons in presence of (light) Majorana neutrinos.
For this purpose we derive a model-independent general 
parametrization of neutrino mass matrices with physically interpretable and irreducible set of parameters.
The parametrization is valid for any number of neutrinos and interpolates smoothly between the heavy Majorana and the (pseudo)Dirac neutrino limits.
We apply the new parametrization to the study of  $P \rightarrow \nu \nu$ and $P \rightarrow \nu \nu \gamma$  decays within the SM extended by additional 
singlet fermions.
We update the SM predictions for the branching ratios of $B_{s,d} \rightarrow \nu \nu \gamma$ and discuss the sensitivity of the $B_{s,d} \rightarrow E_{\rm miss} (\gamma)$ decays
to neutrino mass and mixing parameters.

\end{abstract}

\maketitle


\section{Introduction}\label{intro}


The discovery of neutrino oscillations~\cite{Fukuda:1998mi} 
implies the existence of at least two massive neutrino species. On the other hand, many theoretical models of neutrino mass generation, including the simplest canonical see-saw mechanism~\cite{GellMann:1980vs, Minkowski:1977sc, Mohapatra:1979ia, Magg:1980ut, Lazarides:1980nt},  predict the existence of additional electromagnetically neutral massive fermions.
In general, the neutrino spectrum consists of $3 + n_{N}$ fermions. Three of them 
are the so-far observed standard model (SM) like ($\nu^M_j$) neutrinos. Possible additional $n_{N}$ massive neutrinos ($N^M$)  have not yet been observed. In the following and without loss of generality we assume them to be of Majorana type.\footnote{
The model of Dirac SM-like neutrinos is a special case with $n_{N} = 3$ and with all Majorana mass terms set to zero.
Its spectrum consists of 3 Dirac neutrino fields, which can be written as a superposition of
$\nu^M_j$ and $N^M_j$  fields.}
In the last few decades many different mechanisms have been proposed to explain the smallness of the observed neutrino masses. In the canonical see-saw mechanism for example, heavy $N^M_k$ neutrinos induce 
 small Majorana masses for the observed $\nu^M_j$ neutrinos via their Yukawa interactions.
In this scenario $N^M_k$ neutrinos are typically too heavy to be directly produced in terrestrial experiments~\cite{Ibarra:2010xw}.
On the other hand, in models with approximate lepton number conservation, these new degrees of
freedom could also naturally appear at low energies, see e.g.~\cite{Sierra:2012yy}.
In fact there are several circumstantial motivations for considering additional light $N^M_k$ neutrinos.
Massive neutral fermions which are long lived enough compared to the age of the universe and have mass in the
range  ${2}\rm{keV} \lesssim m_{N_k} \lesssim {5}\rm{keV}$~\cite{Abazajian:2001vt, Viel:2005qj} are good warm dark matter candidates~\cite{Dodelson:1993je,Abazajian:2001nj,Shi:1998km,Dolgov:2000ew}.
Additional $N^M_k$ neutrinos with masses in the range ${1}\rm{GeV} \leq m_{N_k} \ll {100}\rm{GeV}$ are also predicted in models of cosmological Baryon asymmetry generation through neutrino oscillations~\cite{Asaka:2005an, Asaka:2005pn}. 
Finally, persistent tensions in the interpretation of some neutrino oscillation experiments and cosmological observations might imply the existence of additional $N^M_k$ neutrinos with masses at the eV scale, see e.g.~\cite{Abazajian:2012ys}.

An important aspect of neutrino mass model building involves consistently taking into account existing experimental information on low energy neutrino masses and mixings. In principle these inputs can be used to reduce the number of free model parameters. In practice however, this requires a detailed a priori knowledge of how elements of neutrino mass and mixing matrices are connected with each other. In the limit of heavy $N^M_k$ neutrinos such connections are given explicitly by the Casas-Ibara parametrization~\cite{Casas:2001sr}. 
In the last few years, more general parameterizations have been proposed, which do not rely on expansions in small mass ratios and are thus valid away from the limit of heavy $N^M_k$ neutrinos. To date such parametrizations have been found for the case of two~\cite{Donini:2012tt,Bolton:2019pcu} or three~\cite{Blennow:2011vn, Hernandez:2014fha, Agostinho:2017wfs} additional $N^M_k$ neutrinos. Most recently, a master parametrization applicable for the most general case including neutrino mass generation beyond see-saw models has also been proposed~\cite{Cordero-Carrion:2019qtu}. However, the generality comes with several drawbacks: its parameters lack intuitive physical interpretability, the connections between the $N^M_k$ neutrino masses and the SM-like neutrino Yukawa couplings are somewhat obscured. 
In this paper we build upon and extend previous work~\cite{Donini:2012tt,Agostinho:2017wfs} and derive a model-independent general parametrization of the neutrino mass matrices that covers and interpolates between all see-saw like scenarios, including the heavy Majorana mass limit and the pseudo-Dirac case for {\it any number of} additional $N^M_k$ neutrinos (see e.g. Ref.~\cite{Arkani-Hamed:2016rle} for an explicit model realization of such a scenario). Its main purpose is to better map out the possible neutrino mass parameter space and to help create a consistent picture of $N^M_k$ neutrinos at low energies, which is a starting point for developing UV neutrino mass models.\footnote{{We note that in UV models where $B - L$ is gauged (like in $U(1)_{B-L}$~\cite{Yanagida:1979as} or left-right symmetric models~\cite{Minkowski:1977sc, Mohapatra:1979ia}), anomaly cancellation requires exactly three right-handed neutrinos ($n_N=3$).}}

We demonstrate the usefulness of the parametrization using the example of 
$P \rightarrow \nu \nu$ and $P \rightarrow \nu \nu \gamma$ decays, previously studied in Ref.~\cite{Badin:2010uh} in the context of light dark matter searches, 
where $P$ is a neutral pseudoscalar meson and $\nu$ includes both $\nu_j^M$ as well as possibly $N^M_k$ if kinematically allowed. We estimate the contribution of these
two decay topologies to the effective invisible decay widths of neutral mesons (assuming $N^M_k$ are long lived enough to escape the detectors) and show how they are
 sensitive to neutrino mass and mixing parameters. Along the way we also update the theoretical predictions for $B_{s,d} \rightarrow \nu \nu \gamma$ in the SM using state-of-the-art inputs~\cite{Kozachuk:2017mdk} for the relevant hadronic parameters and the associated uncertainties. 
The $P \rightarrow \nu \nu$ decays are helicity suppressed and therefore negligible
in the SM with Dirac neutrinos~\cite{Badin:2010uh} as well as in the limit of heavy $N^M_k$. 
However in models with light $N^M_k$, such that they can appear in the final state, their branching fractions
can become significant. 
On the other hand $P \rightarrow \nu \nu \nu \nu$ decays
are not helicity suppressed~\cite{Bhattacharya:2018msv}, 
however their contributions to the  invisible $P$ decay widths
turn out (within our assumptions) to be completely negligible.
Experimentally, the best sensitivity is expected from $B_{s,d}$ meson decays into unobserved decay products (registered as missing energy $E_{\rm miss}$ in the detector) 
which have already been searched for by the Belle~\cite{Hsu:2012uh} and BaBar~\cite{Lees:2012wv} collaborations. At present the tightest upper limit of $\mathcal{B}r(B_d \rightarrow E_{\rm miss}) < 2.4 \times 10^{-5}$ at 90$\%$ confidence level is provided by BaBar~\cite{Lees:2012wv}. While searches for invisible $B_s$ decays
have not yet been attempted, they are planned at the Belle II experiment, which is also expected to improve significantly the 
upper bound on $\mathcal{B}r(B_d \rightarrow E_{\rm miss})$~\cite{Kou:2018nap}.

The paper is organized as follows. In Sec.~\ref{Sec: parametrization} we derive a
general parametrization of neutrino mass matrices for an arbitrary number of additional massive fermionic singlets and explore the heavy Majorana
neutrino limit and the pseudo-Dirac limit.  We also present the basic properties of the
parametrization and how these can be used to extract neutrino parameters from experiments. 
In Sec.~\ref{Sec: Pinv} we study  the $P \rightarrow \nu \nu$ and 
$P \rightarrow \nu \nu \gamma$ decays separately, estimate their 
contributions to the invisible decay widths of $B_{d,s}$ mesons, and discuss their dependence on the neutrino parameters. 
We summarize our findings in Sec.~\ref{conclusions}. Analytical expressions for the  $P\rightarrow \nu \nu$ and $P \rightarrow \nu \nu \gamma$
decays as well as the details on the perturbative and non-perturbative QCD inputs used in this work are given in 
 Appendix~\ref{Appendix}, while Appendix~\ref{appendixB} contains the details on the derivation of a lower bound on the Frobenius norm of the neutrino mixing matrix.


\section{Parametrizing neutrino masses and mixing in presence of light Majorana neutrinos}
\label{Sec: parametrization}


\subsection{Setup and notation}
\noindent
We consider a family of neutrino models at low energies which are described 
by the Lagrangian\,\footnote{
	We use the formalism presented in Ref.~\cite{Atre:2009rg}.
	A comprehensive discussion of Dirac, Weyl and Majorana fields is given in Ref.~\cite{Pal:2010ih}.}
$\mathcal{L} = \mathcal{L}^{\text{SM}} + \mathcal{L}^{N}$, where
$\mathcal{L}^{\text{SM}}$ is the Standard Model (SM) Lagrangian and 
$ \mathcal{L}^{N}$ is given by:
\begin{equation}\label{eq: Lagrangian}
\begin{split}
\mathcal{L}^{\rm{N}} = -& 
\sum_{a=1}^3 \sum_{b=1}^n
\overline{\nu_{aL}} \thinspace (M_D)_{ab}  N_{bR} -
\frac{1}{2} \thinspace \sum_{b=1}^n \sum_{b'=1}^n
\overline{(N_{bR})^c} \thinspace (M_M)_{bb'}  N_{b'R} +
\mathrm{h.c.}\\[2mm]
+&\sum_{b = 1}^n 
\overline{N_{bR}} \thinspace \mathrm{i}\gamma^{\mu} \partial_{\mu}  N_{bR}\,. 
\end{split}
\end{equation}
The first term is the Dirac mass term where the Yukawa coupling matrix $y$ is implicit in 
$M_D = v/\sqrt{2}y$, where $v$ is the Higgs VEV. 
The second term is the Majorana mass term of chiral right-handed neutrinos $N_{bR}$. 
The mass matrices of models which preserve SM local symmetries  and are renormalizable (chiral left handed neutrino mass terms are forbidden) form a symmetric
block matrix $M$ of the form\footnote{The symmetric nature of $M$ is typical for see-saw like models of neutrino mass generation and is central to our parametrization. For more general scenarios leading to non-symmetric $M$, the parametrization of Ref.~\cite{Cordero-Carrion:2019qtu} applies.}
\begin{equation} \label{eq: full_mass_matrix}
\mathcal{L} \supset- \frac{1}{2}\begin{pmatrix}
\overline{\nu_L} \ \ &
\overline{(N_{R})^c}
\end{pmatrix}
\begin{pmatrix}
0_{3 \times 3} \ \ & (M_D)_{3 \times n} \\[2mm]
(M^T_D)_{n \times 3} \ \ & (M_M)_{n \times n} \\
\end{pmatrix}
\begin{pmatrix}
(\nu_L)^c\\[2mm]
N_R\\
\end{pmatrix}+\mathrm{h.c.}\,.
\end{equation}
The matrix $M$ can be diagonalized by the unitary 
	matrix $L$ in the following way
\begin{align}
M_{\rm{diag}} =
L^\dag 
\begin{pmatrix}
0 \ \ & M_D\\[2mm]
M^T_D \ \ &   M_M\\
\end{pmatrix}
L^*
=
\begin{pmatrix}
D_{\nu} \ \ & 0\\[2mm]
0 \ \ & D_{N}\\
\end{pmatrix},
\quad \quad \quad \text{where} \quad \quad \quad
&L =
\begin{pmatrix}
U_{3 \times 3} \ \  & V_{ 3 \times n}\\[2mm]
X_{ n \times 3} \ \ & Y_{n \times n}\\
\end{pmatrix}.
\end{align}
Here $D_\nu = \text{diag}(m_{\nu_1}, m_{\nu_2}, m_{\nu_3})$  is the diagonal mass matrix 
of SM-like neutrinos ($\nu^M_j = [\nu_{jL} + (\nu_{jL})^c]_m$) and $D_N = \text{diag}(m_{N_1},..., m_{N_n})$ is the diagonal mass
matrix of $N$ neutrinos ($N^M_k = [N_{kR} + (N_{kR})^c]_m$), where $(\nu_{jL})_m$ and $(N_{kR})_m$ are the mass eigenstates which are connected to the gauge interaction eigenstates via
\begin{equation}\label{eq: L}
\begin{pmatrix}
{\nu_L}\\[2mm]
(N_R)^c\\
\end{pmatrix}=
L
\begin{pmatrix}
{\nu_L}\\[2mm]
(N_R)^c\\
\end{pmatrix}_{\rm{m}}.
\end{equation}
Without loss of generality, all diagonal elements of $D_\nu$ and $D_N$ can be taken as real (via suitable unphysical phase rotations of the neutrino fields). 
With this notation one can write down the interaction terms in the Lagrangian
expressed by $\nu^M_j$ and $N^M_k$ neutrino mass eigenstates  and
by $l_{m} \in \{e, \mu, \tau\}$ charged lepton mass eigenstates,

\begin{equation}\label{eq_int}
\begin{split}
\mathcal{L} \supset& - \frac{g W_{\mu}^+}{\sqrt{2}} \sum_{l_m \in \{ e, \mu, \tau \} } 
\left( \sum_{j=1}^3 (U^{\dag}O_L)_{jm} 
\overline{\nu^M_j} \gamma^{\mu} P_L l_m +
\sum_{k=1}^{n} (V^{\dag} O_L)_{km}
\overline{N^M_{k}} \gamma^{\nu} P_L l_m  \right) +\mathrm{h.c.}\\[2mm]
&- \frac{g Z_{\mu}}{2 \cos{\theta_W}}  \left( \sum_{j,j'=1}^3
(U^{\dag}U)_{j j'}\thinspace \overline{\nu^M_{j}} \gamma^{\mu} P_L \nu^M_{j'}+
\sum_{k,k'=1}^{n} (V^{\dag} V)_{kk'}\thinspace
\overline{N^M_{k}} \gamma^{\nu} P_L N^M_{k'} \right)\\[2mm]
&- \frac{g Z_{\mu}}{2 \cos{\theta_W}}   \left(  \sum_{j=1 }^3 \sum_{k=1}^{n} 
(U^{\dag}V)_{jk}\thinspace \overline{\nu^M_{j}} \gamma^{\mu} P_L N^M_{k}
+ \mathrm{h.c.}  \right),
\end{split}
\end{equation}
where $O_L$ and $O_R$
are unitary matrices diagonalizing the mass matrix $M^l$ of the charged leptons  
via the biunitary transformation 
$O^\dag_L M^l O_R = \text{diag}(m_e, m_\mu, m_\tau)$ and $P_{L,R} = (1 \mp \gamma_5)/2$\,.

\subsection{Derivation}

In this section we consider a model with $n_{\nu}$ neutrinos $\nu^M_j$ and 
$n_N$ neutrinos $N^M_k$. 
Matrices $U$, $V$, $X$ and $Y$  clearly depend on the neutrino masses, 
thus it is appropriate to have a simple parametrization of these matrices in terms of
physical neutrino parameters.
Parametrization of this type can be derived with a few simple steps.
We start by decomposing the $V$ matrix into 
\begin{equation}
V = \left\{
\begin{array}{lll}
g  S &:& n_{N} > n_{\nu},\\[2mm]
g  P &:& n_{N} \leq n_{\nu},
\end{array}
\right.
\end{equation}
where $g$ is a $n_{\nu} \times n_{\nu}$ complex matrix, $S$ is given by
$S_{n_{\nu} \times n_N} = [ I_{n_{\nu} \times n_{\nu}}, \
\hat{S}_{n_{\nu} \times (n_N - n_{\nu})} ]$ and $P_{n_\nu \times n_N}$ 
is a {\it projection} matrix which in the case of a normal hierarchy of $\nu^M_j$ neutrino masses is given by 
\begin{equation}
P = \begin{pmatrix}
\ \  0_{(n_\nu - n_N) \times n_N}\\[2mm]
I_{n_N \times n_N}\\
\end{pmatrix}.
\end{equation}
The $\hat{S}$ matrix is a general complex $n_{\nu} \times (n_N - n_{\nu})$ matrix.
Its elements can be chosen freely.
By using the definition of the transition matrix, $L M_{\rm{diag}} L^T =M $ 
one obtains the relation
\begin{align}
U D_\nu U^T + V D_N V^T = 0\,,
\end{align}
which, depending on $n_N$, can be rewritten as
\begin{subequations}
\begin{align}
\label{eq: UDU1}
U D^{1/2}_\nu D^{1/2}_\nu U^T &= g (- S D_N S^T)^{1/2}(- S D_N S^T)^{1/2} g^T
&: n_N > n_\nu\,,	& & \vspace{2mm} \\[2mm] 
\label{eq: UDU2}
U D_{\nu} U^T &= 
g P (- D_N)^{1/2}(- D_N )^{1/2} P^T g^T
&: n_N \leq n_\nu\,. & &
\end{align}
\end{subequations}
Eq.~(\ref{eq: UDU1}) is already in the desired form. With a few assumptions it can
be written as $RR^T = I$, where $R$ is an invertible orthogonal complex squared matrix, which
connects $U D^{1/2}_\nu$ with $g (- S D_N S^T)^{1/2}$.
On the other hand Eq.~(\ref{eq: UDU2})  is not yet of the desired form.
In the special case $n_{N} = n_\nu$ the projection matrix $P$ becomes the identity matrix,
and we can decompose the left-hand side of Eq.~(\ref{eq: UDU2}) into
$U D_{\nu} U^T = U D^{1/2}_\nu D^{1/2}_\nu U^T$.  Notice that on both sides
all matrices have the same shape.
In the case $n_{N} < n_{\nu}$ the diagonal mass matrix $D_{\nu}$ is not invertible.
This can be seen directly by looking at the block mass matrix $M$,
\begin{equation}
M = \begin{pmatrix}
0_{n_{\nu} \times n_{\nu}} \ \ & (M_D)_{n_{\nu} \times n_N} \\[2mm]
(M^T_D)_{n_N \times n_{\nu}} \ \ & (M_M)_{n_N \times n_N} \\
\end{pmatrix}.
\end{equation}
Of the first $n_{\nu}$ rows, maximally $n_N$ are independent. Similarly,
of the last $n_N$ rows,  also maximally $n_N$ are independent. Therefore
one can construct at least $n_{\nu} + n_N - 2 n_N = n_{\nu} - n_N$  
{ independent} eigenvectors
for the matrix $M$ with zero eigenvalues. From here, there are { maximally} $n_N$ non zero 
masses $m_{\nu_j}$ in $D_\nu$. 
This property can be written as
\begin{equation}
\label{eq: prop}
D_{\nu} = P (P^T D_\nu P) P^T,  
\end{equation}
where $P^T D_\nu P$ is a $n_N \times n_N$ diagonal matrix of all non zero eigenvalues of $D_\nu$, therefore an invertible positive definite matrix. 
By specifying $D_\nu$ one can use this property to find $P$.
Using Eq.~(\ref{eq: prop}) one can finally rewrite the left-hand side of 
Eq.~(\ref{eq: UDU2}) in terms of $n_N \times n_N$ matrices,
\begin{equation}
\label{eq: UDU_P}
(\tilde{P}^T U P) (P^T D^{1/2}_\nu P)(P^T D^{1/2}_\nu P)( P^T U \tilde{P}) = 
(\tilde{P}^T g P) (- D_N)^{1/2}(- D_N )^{1/2} (P^T g^T \tilde{P})\,.
\end{equation}
Here we have multiplied the equation by a $n_{\nu} \times n_N$ arbitrary 
 matrix $\tilde{P}$ from the right-hand side and by $\tilde{P}^T$ from the left-hand side. 

At this point we assume that all matrices, which are products of matrices inside brackets
in Eq.~(\ref{eq: UDU_P}) and  Eq.~(\ref{eq: UDU1}), are invertible, 
$D_N$ matrix is invertible for all pairs $(n_\nu, n_N)$ and that 
matrices $U$ and $g$ are also invertible in the case $n_\nu < n_N$.
Cases in which these assumptions do not hold can still be handled by the parametrization
we are deriving, by taking appropriate limits.
Within our assumptions, both Eqs. (\ref{eq: UDU1}) and (\ref{eq: UDU_P})
can be expressed in the form $R^T R = I$ or $R R^T = I$ equivalently, 
where $R$ is a $\text{min}( n_\nu \times n_\nu, \ n_N \times n_N )$ 
general complex orthogonal matrix. It links $U$ and $V$ matrices through
\begin{equation}
\label{eq: vuq}
V =  UQ\,,
\end{equation}
where the $ n_\nu \times n_N$ matrix $Q$ is given by
\begin{equation}
Q = \left\{
\begin{array}{lll}
 D^{1/2}_{\nu} R (-S D_N S^T)^{-1/2} \ S &:& n_N > n_\nu\,,\\[2mm]
 D^{1/2}_{\nu} PR (-D_N)^{-1/2} &:& n_N \leq n_\nu\,.\\
\end{array}
\right.
\end{equation}
In case $n_\nu > n_N$ one obtains the relation $\tilde{P}^T(V-UQ) = 0$ which leads
to $V = UQ$, since $\tilde{P}^T$ is arbitrary up to the above assumptions.
Eq.~(\ref{eq: vuq}) together with the unitarity 
condition $LL^\dag = I$
finally leads to the desired parametrization of $U$ and $V$ matrices
\begin{subequations}
\begin{align}
UU^\dag + VV^\dag =& \ I,\\
U\sqrt{  I + QQ^\dag}\sqrt{  I + QQ^\dag} U^\dag =& \ I,\\
\label{eq: U}
U =& \ A \left( I + QQ^\dag \right)^{-1/2}.
\end{align}
\end{subequations}
Here $A$ is a $n_\nu \times n_\nu$  unitary matrix. 
The $I + QQ^\dag$ hermitian matrix is positive definite, which means
that it has precisely one positive definite square root which is also invertible.
Parametrizations of $X$ and $Y$ matrices
are similarly obtained from the unitary condition $LL^\dag = I$ using 
the above derived results.

\subsection{Main formulae}

Below we give the full set of equations which define the parametrization 
of the neutrino mixing matrices:
\begin{subequations}
\begin{align}
\label{eq_U}
U&=A \left( \sqrt{I + QQ^\dag} \right)^{-1},\\[1mm]
\label{eq_V}
V&=A  \left( \sqrt{I + QQ^\dag} \right)^{-1}Q,\\[1mm]
\vspace{2mm}
\label{eq_X}
X&=-B \left(\sqrt{I + Q^\dag Q}\right)^{-1}Q^\dag,\\[1mm]
\label{eq_Y}
Y&=B \left(\sqrt{I + Q^\dag Q}\right)^{-1},
\end{align}
\end{subequations}
where $A_{n_\nu \times n_{\nu}}$ and $B_{n_N \times n_N}$ are unitary matrices and Q is defined by
\begin{equation}
Q = \left\{
\begin{array}{lll}
- \mathrm{i} D^{1/2}_{\nu} R (S D_N S^T)^{-1/2} \ S &:& n_N > n_\nu,\\[2mm]
- \mathrm{i} D^{1/2}_{\nu} PR D^{-1/2}_N &:& n_N \leq n_\nu.\\
\end{array}
\right.
\end{equation}
The { $M_D =  v/\sqrt{2} y $} and $M_M$ matrices are therefore parametrized via
\begin{subequations}
\begin{align}
\label{eq_M_D}
M_D =& \ U D_\nu X^T +  V D_N Y^T,\\[1mm]
\label{eq_M_M}
M_M =& \ X D_\nu X^T + Y D_N Y^T ,
\end{align}
\end{subequations}
where $D_\nu$ and $D_N$ are diagonal neutrino mass matrices 
$D_\nu = \text{diag}(m_{\nu_1},...,  m_{\nu_{n_\nu}})$ and
$D_N = \text{diag}(m_{N_1},...,  m_{N_{N_n}})$ defined up to arbitrary phases
(for each diagonal element).
The derived parametrization expresses neutrino mixing matrices appearing in the Lagrangian 
in terms of unitary matrices $A$ and $B$, complex orthogonal matrix $R$, 
neutrino masses contained in  $D_\nu$ and $D_N$ and
in case $n_N > n_\nu$ also a general complex matrix 
$\hat{S}$ which is hidden inside 
$S = [I_{n_{\nu} \times n_{\nu}}, \hat{S}_{n_{\nu} \times (n_N-n_{\nu})}]$.
The { projector} $P$ in case $n_\nu \geq n_N$ must be chosen such that
 $P^T D_\nu P$ is a diagonal matrix of all non zero eigenvalues of $D_\nu$
and that the relation $P P^T D_\nu P P^T = D_\nu$ holds. If $n_\nu > n_N$,
the choice of $P$ depends on the hierarchy of $\nu^M_j$ neutrino masses.

\subsection{Physical case $n_{\nu} = 3$}

In the following we discuss the explicit form of our parametrization, it's limits and parameter counting, for the realistic case $n_\nu=3$ and various possible choices of $n_N$\,.

In the case $n_{\nu} = n_N = 3 $ previously studied in Ref.~\cite{Agostinho:2017wfs} the $R$ matrix can be parametrized with 3
complex angles:
\begin{equation}\label{R}
R =
\begin{pmatrix}
c_1 \ \ &  \pm s_1 \ \ & 0\\[1mm]
-s_1\ \ &  \pm c_1 \ \ & 0\\[1mm]
0 \  \  &      0       \ \ & 1\\
\end{pmatrix}
\begin{pmatrix}
c_2 \ \ & 0 \ \ & \pm s_2\\[1mm]
0    \ \ & 1 \ \ & 0          \\[1mm]
-s_2\ \ & 0 \  \ & \pm c_2\\
\end{pmatrix}  
\begin{pmatrix}
1 \ \ & 0     \ \ &  0           \\[1mm]
0 \ \ & c_3  \ \ & \pm  s_3 \\[1mm]
0 \ \ & -s_3 \ \ & \pm c_3  \\
\end{pmatrix},
\end{equation}
where $c_j = \cos( \phi_j + \mathrm{i} \theta_j )$ and
$s_j = \sin( \phi_j + \mathrm{i} \theta_j )$ with $\phi_j \in [0, 2 \pi) $ and
 $\theta_j \in {\rm I\!R}$ for $j \in \{ 1,2,3 \}$.
Free signs in each of the three matrices must be equal 
(both  $+$ or  both $-$ in each matrix). 
The { projector} $P$ in this case reduces to the identity matrix.

In the case $n_\nu=3$ with $n_N = 2$ previously studied in Ref.~\cite{Donini:2012tt} one of the $\nu^M_j$ neutrinos is massless, 
thus the $D_\nu$ matrix can be parametrized 
by $D_\nu = \text{diag}(0, m_{\nu_2}, m_{\nu_3})$ in case of normal hierarchy (NH) or 
by $D_\nu = \text{diag}(m_{\nu_1}, 0,  m_{\nu_3})$ in case of inverted hierarchy (IH). 
In both scenarios the $R$ matrix is described by one complex angle:
\begin{equation}
R =\begin{pmatrix}
\ \ \cos(\phi+ \mathrm{i} \theta)& \ \ \pm \sin(\phi+ \mathrm{i} \theta)\\[2mm]
   -\sin(\phi+ \mathrm{i} \theta)& \ \ \pm \cos(\phi+ \mathrm{i} \theta)\\
\end{pmatrix},
\end{equation}
where $\phi \in [0, 2 \pi)$ and  $\theta \in {\rm I\!R}$. As before the free signs in the $R$  matrix must be equal (both  $+$ or  both $-$).
The  projector $P$ for NH (IH) is given by:
\begin{equation}
P^{\mathrm{NH}} =\begin{pmatrix}
0\ \ & 0 \\[1mm]
1\ \ & 0 \\[1mm]
0\ \ & 1 \\
\end{pmatrix} \quad \quad \text{and} \quad \quad
P^{\mathrm{IH}} =\begin{pmatrix}
1\ \ & 0  \\[1mm]
0\ \ & 1  \\[1mm]
0\ \ & 0  \\
\end{pmatrix}.
\end{equation}
Also, in the case $n_\nu > n_N$ one can rename the $PR$ matrix to $R_{n_\nu \times n_N}$, since the projector $P$ is always multiplied by $R$ from the right-hand side.

On the other hand, matrix $B$ is a general $n_N \times n_N$ unitary matrix for all pairs $(n_\nu, n_N)$, 
while matrix $A$ is a general $n_\nu \times n_\nu$ unitary matrix only if $n_\nu \leq n_N$,
since if $n_\nu > n_N$ not all generators of U($n_\nu$) are required to form $A$ so that
the $(M_D)_{n_{\nu} \times n_N}$ and $(M_M)_{n_N \times n_N}$ matrices 
are fully parametrized with all $n^2_N + n_N (2 n_\nu +1)$ parameters.
The number of real parameters for each matrix which appears in the derived parametrization 
for the case $n_{\nu} = 3$ is given in Table~\ref{table: d.o.f.}.
\begin{table}[h]
	\centering
	\caption{Number of parameters in the derived parametrization of the neutrino mass matrices of the extended SM with $n_\nu = 3$ and $n_N = n$ additional chiral right-handed neutrinos.}
	\setlength{\tabcolsep}{6pt}
	\renewcommand{\arraystretch}{1.5}
	\begin{tabular}{ c | c  c  c  c  c  c |rl}
		\hline \hline
			$n$	 &  \ $ (D_\nu)_{3 \times 3} $ \ & 
		\ $(D_N)_{n \times n}$  \  & 
		\ $R_{\text{min}(n \times n, \ 3 \times 3)}$  \ & 
		\ $\hat{S}_{3 \times (n-3)}$ \ & 
		\ $A_{3\times 3}$ \ &
		\ $B_{n\times n}$ \ & 
		\ total $n^2 +7n$\ \\   \hline 
		$n=1$ & 1 & 1     & 0 &      /        & { 5} &      1     &   8  \\ 
		$n=2$ & 2 & 2     & 2 &      /        & {8} &      4     & 18  \\ 
		$n=3$ & 3 & 3     & 6 &      /        & 		    9 	   	    &      9      & 30  \\ 
		$n>3$ & 3 & $n$ & 6 & $6(n-3)$ & 			9			 & $n^2$  & $n^2 +7n$           \\ \hline \hline
	\end{tabular}
	\label{table: d.o.f.}
\end{table}
{Note that interactions between neutrinos (both $\nu_j$ and $N_k$) and other SM particles do not depend on the $B_{n \times n}$ matrix. }

 \subsection{Heavy $N_k$ limit}
 
 In the limit where $|| D_{\nu}|| / || D_{N}|| \ll 1$ and $||Q|| \ll 1$ the parametrization
  simplifies as $U \approx A$, $V \approx AQ$, $X \approx -BQ^\dag$, $Y = B$ resulting in
 \begin{equation}
M_D = \left\{ 
\begin{array}{ll}
\ - \mathrm{i} A D^{1/2}_\nu R \ (S D_N S^T)^{-1/2} \ S D_N B^T &: n_N > n_\nu\,, \\[2mm]
\quad 
 \mathrm{i} A D^{1/2}_\nu P R \ D^{1/2}_N B^T &: n_N \leq n_\nu\,,
\end{array}
\right.
\end{equation}
and
 \begin{align}
 M_M \approx   B D_N  B^T\,.
 \end{align}
In case $n_N \leq n_\nu$ we immediately recognize the original Casas-Ibara parametrization~\cite{Casas:2001sr}. 
By combining the expressions of $M_D$ and $M_M$ matrices one can also construct the well known effective $\nu^M_j$ neutrino mass matrix
\begin{equation}
M^{\rm{effective}}_{\nu} \equiv -M_D M^{-1}_M M^T_D \approx A D_\nu A^{T}\,.
\end{equation}
This formula is valid for all pairs $(n_\nu, n_N)$.
In this scenario the PMNS matrix which is given by 
$U_{\text{PMNS}} = O^\dag_L U \approx O^\dag_L A$ (see Eq.~(\ref{eq_int})) 
is approximately unitary. 
To clarify, the PMNS matrix maps fields of observed neutrinos from the
mass basis into the flavor (gauge) basis. In the heavy neutrino limit low energy processes can only involve $\nu^M_j$ neutrinos.

 \subsection{Dirac neutrino limit}
\label{Dirac limit}

In the scenario with $n \equiv n_\nu = n_N$ and $M_M = 0$ neutrinos can be described
by Dirac fields (linear combinations of Majorana fields which are not Majorana fields). 
The condition $M_M = 0$  is equivalent to $R^\dag |D_\nu|^2 R^* = |D_N|^2$.
The phases in front of neutrino masses are arbitrary and do not affect any
measurable quantities. Therefore we can fix the phases by choosing 
$D \equiv D_\nu = D_N$, where $D$ is a positive definite mass matrix.
From here, one finds   $Q = -\mathrm{i} R$, where
$R = \text{diag}(\pm 1,..., \pm 1)$. Signs are arbitrary and independent 
of each other. The transition matrix $L$ then takes the following form
\begin{align}
L = \frac{1}{\sqrt{2}}
\begin{pmatrix}
A \ \                       &  -\mathrm{i}AR\\[2mm]
-\mathrm{i} BR \ \  & B                      \\ 
\end{pmatrix}.
\end{align}
At this point one can diagonalize $M_D$ through a biunitary transformation
\begin{align}
M_D = - \mathrm{i} ARDB^T\,.
\end{align}
Therefore the mass term in the Lagrangian can be written as
\begin{align}
\mathcal{L} \supset - \overline{\nu}^D D \nu^D\,,
\end{align}
\noindent
where
\begin{equation}
\label{eq: D_fields}
\nu^D =  \frac{1}{\sqrt{2}} \left( i R  \nu^M_j + N^M_j \right)\,,
\end{equation}
are Dirac neutrino fields (not Majorana fields).
One can use this equation to {\it define} $\nu^M_j$ and $N^M_j$
Majorana fields starting from the Dirac field. Such definition of Majorana
fields can then be applied outside the Dirac limit. In this way, the obtained model is fully
consistent with both the general Majorana neutrino model (outside the Dirac limit) as well as with the Dirac neutrino model (in the Dirac limit). We use this model in
Section \ref{Sec: Pinv}. 
Finally, the Lagrangian (for the case $n=3$) can be written as
\begin{equation}
\begin{split}
\mathcal{L} \supset \ \ \
& \sum^3_{k=1} \ \overline{\nu^D_k} \left( \mathrm{i} \gamma^{\mu} \partial_\mu  -D \right) \nu^D_k +\\[1mm] 
&+\frac{g}{2 \cos(\theta_W)} Z_{\mu}  \sum_{k=1}^3 \thinspace
\overline{\nu^D_k} \gamma^{\mu} P_L \nu^D_k +\\[1mm]
&+ \frac{g}{\sqrt{2}} W_{\mu}^+ 
\sum_{l_m \in \{ e, \mu, \tau \} } \sum_{k=1}^3
(A^{\dag} O_L)_{km} \
\overline{\nu^D_k}  \gamma^{\mu}P_L l_m +\text{h.c.}\,.
\end{split}
\end{equation}
From the last equation one can recognize the  (unitary) PMNS matrix as  
$U_{\text{PMNS}} = O^\dag_L A = \sqrt{2} O^\dag_L U$.
Its matrix elements are precisely the same as in the heavy neutrino limit if
we keep $A$ matrix unchanged. However, now the observed neutrinos
are 3 Dirac fermions which are specific linear combinations 
of $\nu^M_j$ and $N^M_j$ fields. Note that the measured values of the PMNS matrix can be directly related to the underlying neutrino
parameters only in the heavy neutrino and (pseudo-)Dirac limits.
Outside these two limits the precise relations are non-trivial, since the PMNS matrix
is no longer unitary. However, results from experiments 
which measure the PMNS matrix elements can still be used to constrain parameters 
of the general low energy neutrino models. 

\subsection{Scaling relations}
\label{subsec scaling}

Matrices $U^\dag U$, $U^\dag V$ and $V^\dag V$ satisfy the relation:
\begin{equation}
\label{eq: uu prop}
||U^\dag U||^2 + 2 ||U^\dag V||^2 + ||V^\dag V||^2 = n_\nu\,,
\end{equation}
where $|| U^\dag V ||^2 = \sum_{jj'} | (U^\dag V)_{jj'} |^2 $ is the Frobenius norm
and $n_\nu$ is the number of $\nu^M_j$ neutrinos.
The property can be derived by a few simple steps.
First define matrix ${\mathcal U}$ as
%
\begin{equation}
\mathcal U=\begin{pmatrix}
U^\dag U \ \ & U^\dag V\\[2mm]
V^\dag U \ \ &  V^\dag V\\
\end{pmatrix}.
\label{eq: UCD}
\end{equation}
%
Its relevant properties are ${\mathcal U}^\dag = {\mathcal U}$ and ${\mathcal U}^2 = {\mathcal U}$. The last one holds
due to the unitarity condition $U U^\dag +V V^\dag = I_{n_{\nu} \times n_{\nu}}$.
From here, one gets $\sum_{E=1}^{n_\nu + n_N} {\mathcal U}_{CE} {\mathcal U}_{ED} ={\mathcal U}_{CD}$.
Therefore
%
\begin{equation}
\label{eq: SumUcd}
\sum_{C,D=1}^{n_{\nu} +n_N} |{\mathcal U}_{CD}|^2 =\sum_{C,D = 1}^{n_{\nu}+n_N}
{\mathcal U}_{CD} {\mathcal U}_{DC} = 
\mathrm{Tr}\{ {\mathcal U} \} = n_{\nu}\,,
\end{equation}
%
where the identity $ \mathrm{Tr}\{ V^\dag V \} = \mathrm{Tr}\{ V V^\dag \} $
was used.  
By the definition of the Frobenius norm, relations like $| (U^\dag V)_{jk}| < || U^\dag V ||$
hold. These can be used to constrain matrix elements especially in scenarios
with nearly degenerated $N^M_k$ neutrino masses. {In the heavy Majorana mass limit $m_\nu / m_N \ll 1$ Frobenius norms can be used as direct measures of non-unitarity in the $3\times 3$ light neutrino sector, complementary to the PMNS matrix ($U^\dagger O_L$), since in this limit exact $3\times 3$ unitarity implies $|| U^\dag U || = n_\nu$, while $|| U^\dag V || = || V^\dag V || =0$\,.}

\begin{figure}[h]
	\includegraphics[scale=1.0]{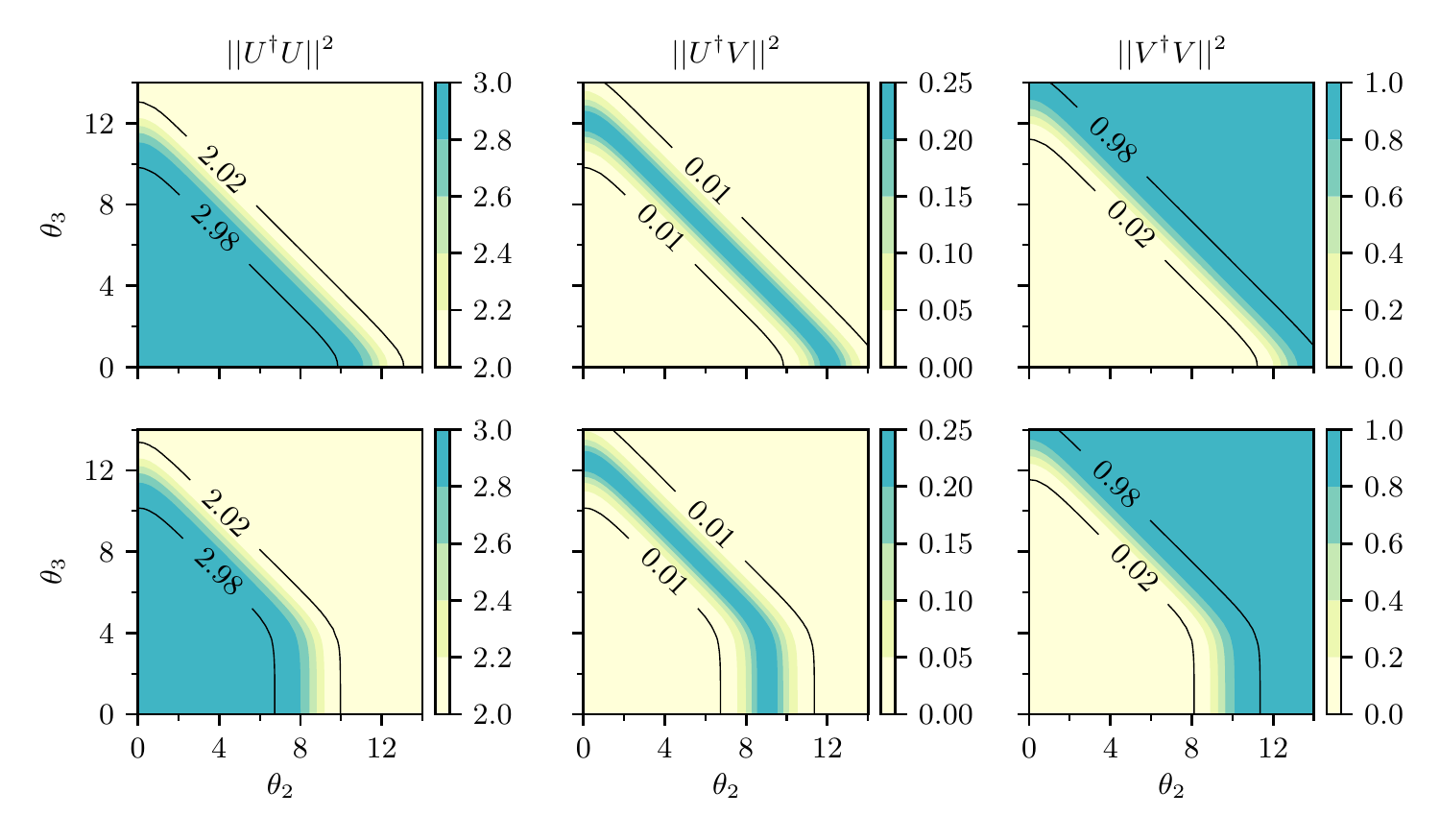}
	\caption{Examples of the $|| U^\dag U ||$, $||U^\dag V||$ and $||V^\dag V||$ 
		norms as functions of the  $\theta_2$ and $\theta_3$ parameters in a model with $n_\nu = n_N = 3$. The other model parameters are fixed as $\theta_1 =0$ and
		$\phi_k = 0$ for all $k$ (implying $\theta_k \rightarrow -\theta_k$ 
		for each $k$). Light neutrino masses are set to satisfy present
		experimental constraints with normal mass hierarchy~\cite{deSalas:2017kay} and $m_{\nu_2}/m_{\nu_1} = 2$\,. 
		The $N^M_k$ neutrino masses are set to 
		$m_N = 1\,\text{GeV}$ (upper plots)
		 and
		$m_{N_1} = 1\,\text{MeV}$, $m_{N_2} = 1\,\text{GeV}$ and 
		$m_{N_3} = 100\,\text{GeV}$ (lower plots).}
	\label{fig: Norms}
\end{figure}

\begin{figure}[h]
	\includegraphics[scale=1.0]{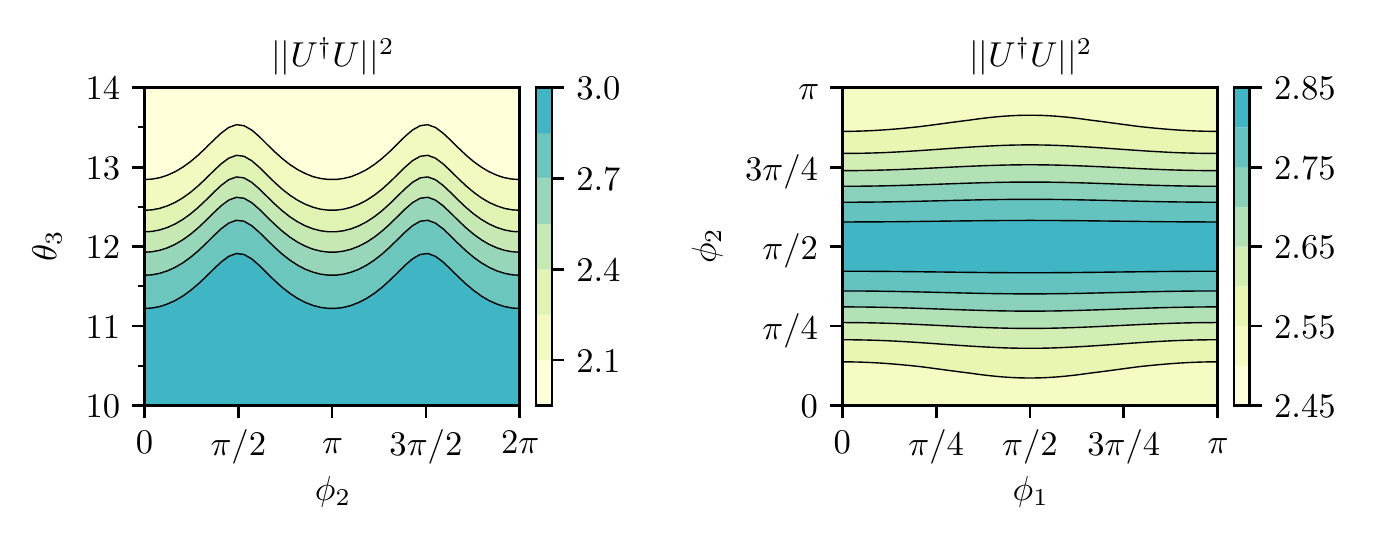}
	\caption{The $|| U^\dag U ||$ norm as a function of $\theta_3$ and $\phi_2$ at $\phi_1=0$ (left), as well as 
		$\phi_2$ and
		 $\phi_1$ at $\theta_3 = 12$ (right), in a $n_\nu = n_N = 3$ scenario with $m_{N_1} = 1\,\text{MeV}$, 
		$m_{N_2} = 1\,\text{GeV}$ and $m_{N_3} = 100\,\text{GeV}$. 
		Other $\theta_j$ and $\phi_j$ parameters are set to zero. Light neutrino masses are set to satisfy present
		experimental constraints with normal mass hierarchy~\cite{deSalas:2017kay} and $m_{\nu_2}/m_{\nu_1} = 2$\,.
	 }
	\label{fig: NormUU_theta_phi}
\end{figure}

{ 
In general $||UU^\dag||$, $||UV^\dag||$ and $||VV^\dag||$ norms can be expressed as a function of
eigenvalues $\mu_j$ of the $QQ^\dag$ matrix:

\begin{align}
|| U^\dag U ||^2 = \sum^{n_\nu}_{j = 1} \frac{1}{(1 + \mu_j)^2}, \quad \quad \quad
|| U^\dag V ||^2 = || 	V^\dag U ||^2 = \sum^{n_\nu}_{j = 1} 
\frac{\mu_j}{(1 + \mu_j)^2}, \quad \quad \quad
|| V^\dag V ||^2  = \sum^{n_\nu}_{j = 1} \frac{\mu^2_j}{(1 + \mu_j)^2},
\end{align}

\noindent
{where the first identity can be related to the determinant of the PMNS matrix defined in the $m_\nu / m_N \ll 1$ limit since $\det(U^\dag U) = \prod^{n_\nu}_{j = 1}{(1 + \mu_j)^{-1}}$.} Number of $\nu^M_j$ neutrinos is also number of required quantities to completely
describe norms. 

In a special case in which $n_\nu = n_N = 3$ and neutrino masses are degenerated,
$D_\nu = m_\nu I$ and $D_N = m_N I$, equations above can be simplified,

\begin{align}
|| U^\dag U ||^2 =& \frac{1}{(1 + \frac{m_\nu}{m_N})^2} +
\frac{1}{(1 + \frac{m_\nu}{m_N} \lambda)^2} + \frac{1}{(1 + \frac{m_\nu}{m_N} \frac{1}{\lambda})^2},\\[2mm]
|| U^\dag V ||^2 =&   \frac{m_\nu}{m_N} \left(
\frac{ 1 }{(1 + \frac{m_\nu}{m_N})^2} +
\frac{ \lambda}{(1 + \frac{m_\nu}{m_N} \lambda)^2} + 
\frac{ \frac{1}{\lambda}}{(1 + \frac{m_\nu}{m_N} \frac{1}{\lambda})^2} \right),\\[2mm]
|| V^\dag V ||^2  =&  \left( \frac{m_\nu}{m_N} \right)^2 \left( 
\frac{ 1 }{(1 + \frac{m_\nu}{m_N})^2} +
\frac{ \lambda^2 }{(1 + \frac{m_\nu}{m_N} \lambda)^2} + 
\frac{  \frac{1}{\lambda^2} }{(1 + \frac{m_\nu}{m_N} \frac{1}{\lambda})^2} \right),
\end{align}

\noindent
where $\lambda$ is the largest eigenvalue of the $R^\dag R$ matrix.
Properties used to derive these equations are described in Appendix \ref{appendixB}.
In this special case norms depend only on $\lambda \in [1, \infty)$ and $m_\nu/m_N$ ratio.
However if $m_\nu/m_N \ll 1$ equations above can be reduced to:

\begin{align}
\label{eq: norms_xi}
|| U^\dag U ||^2 \approx \frac{1}{(1 + \xi^2)^2}+2, \quad \quad \quad
|| U^\dag V ||^2 \approx \frac{\xi^2}{(1 + \xi^2)^2}, \quad \quad \quad
|| V^\dag V ||^2 \approx \frac{\xi^4}{(1 + \xi^2)^2},
\end{align}

\noindent
where $\xi = \sqrt{m_\nu/m_N} \sqrt{\lambda}$. Expressions in Eq. (\ref{eq: norms_xi}) can be regarded as scale relations for norms with respect to a dimensionless scale parameter $\xi$. {Again the first identity is closely related to $\det(U^\dag U) \approx 1/{(1 + \xi^2 )}$ measuring PMNS matrix unitarity.}
If $R$ matrix depends only on one $\theta$ angle, we find $\lambda = \exp(2|\theta|)$. 
In other cases $\lambda$ depends on all parameters of $R$ matrix,
however if $R$ matrix is parametrized in one of specific ways, $\lambda$ depends only on
$\theta_k$ angles and not on $\phi_k$ angles. More details in Appendix \ref{appendixB}.
Dependence of norms with respect to $\theta_k$ and $\phi_k$ parameters are shown in 
Figure \ref{fig: Norms}
and in Figure \ref{fig: NormUU_theta_phi}.
}

\subsection{Extracting neutrino parameters}

Elements of the  $O^\dag_L U$ matrix contain information about
$N^M_k$ and $\nu^M_j$ neutrino masses, $R$ matrix and $S$ matrix elements.
In the case $n_\nu = n_N$ with fixed $D_\nu$ one can simply obtain $R$ and $D_N$ by
properly diagonalizing the left-hand side of
\begin{align}
D^{-1/2}_{\nu} \left[ (U^\dag U)^{-1} - I \right] D^{-1/2}_{\nu} =   
R (-D_N)^{-1} R^\dag \,,
\end{align}
where $U^\dag U = (O^\dag_L U)^\dag  (O^\dag_L U)$, since $O_L$ is unitary. 
In case $n_\nu > n_N$ projectors $P$ on both sides should be added, but if
$n_N>n_\nu$ the diagonalization becomes meaningless, since the $S$ matrix appears 
in various places on the right-hand side of the equation.

Processes involving neutrinos, but not charged leptons in the initial and final state 
usually do not depend on matrices $O^\dag_L U$ and $O^\dag_L V$, 
since masses of charged leptons (in loops) are small compared to the masses
 of $W$ and $Z$ bosons.
Therefore observables in these processes depend only on
neutrino masses, $R$ matrix elements and if $n_{N} > n_{\nu}$ also on
$\tilde{S}$ matrix elements.  
In the heavy $N_k$ and pseudo-Dirac neutrino limits such processes are sensitive to (small) non-unitary corrections to the PMNS matrix, 
but not to PMNS matrix elements themselves. 
As explained above, outside of both limits the precise relation between the experimentally measured PMNS matrix elements and neutrino parameters is highly nontrivial and thus more difficult to interpret.


\section{Neutral meson decays to two neutrinos and (un)resolved photons}
\label{Sec: Pinv}


In this section we consider the decays of pseudoscalar mesons ($P$) to neutrinos and possibly additional photons ($\gamma$),
as prospective venues to constrain 
low energy neutrino parameters.  We consider both signatures of $P \rightarrow E_{\rm miss}$ as well as $P \rightarrow  E_{\rm miss}  \gamma$, where $E_{\rm miss}$ is an energy imbalance registered in the particle detector.
Within our theoretical setup and assuming a $4\pi$ detector coverage with finite EM energy resolution, the decay products which contribute to $P \rightarrow E_{\rm miss}$ include stable enough neutrinos as well as unresolved (soft) photons
\begin{equation}
\label{eq: br1}
\mathcal{B}r(P \rightarrow E_{\rm miss}) =
\mathcal{B}r(P \rightarrow \nu \nu) +
\mathcal{B}r(P \rightarrow \nu \nu   \gamma^* )+
\mathcal{B}r(P \rightarrow \nu \nu \nu \nu )+...\,,
\end{equation}
where the dots denote additional multibody decay channels which are further suppressed. The energy of photons present in the final state ($\gamma^*$), 
should be less than the energy resolution $E_0$ of the EM detector.  According to Refs.~\cite{1742-6596-587-1-012045} and \cite{1742-6596-928-1-012021}
  the EM calorimeter of Belle II
 has the  energy resolution in the range of $20 - 50\, \text{MeV}$.
 For concreteness, we use the value $E_0 = 50 \text{ MeV}$ throughout this paper. 
Light neutrinos are clearly invisible to the detector, while heavy neutrinos may decay
into lighter and observable decay products before they escape.
Thus in general non-trivial conditions which depend on neutrino parameters are imposed 
on the branching ratio $\mathcal{B}r(P \to E_{\rm miss} (\gamma))$.
We aim to study this dependence and in particular to estimate theoretical upper bounds on $\mathcal{B}r(P \rightarrow E_{\rm miss} (\gamma))$
based on the relevant experimental constraints. 
To do so, we first briefly discuss the basic properties of
$P \rightarrow \nu \nu$ and $P \rightarrow \nu \nu \gamma$ decays
in the following framework.\footnote{As we discuss in detail in Sec.~\ref{sec:Pnunugamma}, the $P\to \nu\nu\nu\nu$ contributions are always negligible in our framework.}
We use the Majorana neutrino model with $n_{\nu} = n_{N} = 3$.
Details about the model and all relevant equations to reproduce the results
are described in Appendix~\ref{Appendix}.  We use the following compact notation
\begin{align}
\nu_C = \left\{
\begin{array}{lll}
\nu^M_C      &: &C \in \{1,2,3\}\,,\\[2mm]
N^M_{C-3} &: &C \in \{4,...,3+n\}\,,
\end{array}
\right.
\end{align}
%
along with the $\mathcal{U}$ matrix
to estimate the relevant branching ratios.\footnote{We note in passing that observables which do not depend on $U^\dag O_L$ and
$V^\dag O_L$ matrix elements can contain $\mathcal{U}_{CD}$ in two forms:
$|\mathcal{U}_{CD}|^2$ and/or $\text{Re}( \mathcal{U}^2_{CD})$. 
Both are unaffected by the change 
$\mathcal{U}_{CD} \rightarrow \mathcal{U}^*_{CD}$. 
In the case $n_N \leq n_\nu$, this implies a symmetry $\theta \rightarrow - \theta$,
since $R(\phi_1, \phi_2, \phi_3, \theta_1, \theta_2, \theta_3)^* = 
R(\phi_1, \phi_2, \phi_3, -\theta_1, -\theta_2, -\theta_3)$. Exchanging $\theta_k$ with $-\theta_k$ for a single $k$ is in general not a 
symmetry of observables.}
 Numerical results and plots in this section are calculated using expressions and numerical inputs
described in Appendix \ref{Appendix}.
For light neutrino masses $m_{\nu_j}$ we impose experimental constraints
from \cite{deSalas:2017kay}.  For concreteness,  in cases where observables significantly depend on light neutrino
masses, we assume a normal mass hierarchy with $m_{\nu_2}/m_{\nu_1} = 2$\,.

\subsection{$P \rightarrow \nu \nu$}

\begin{figure}[h]
	\includegraphics[scale=1.0]{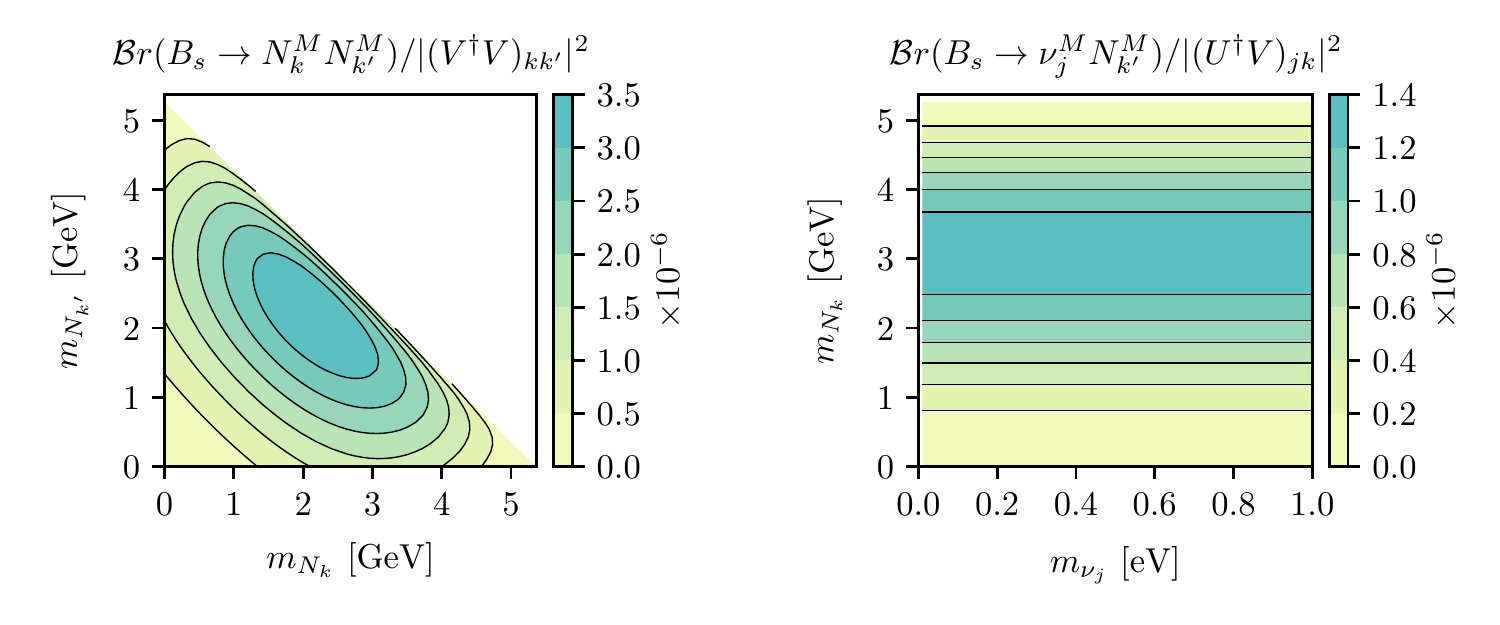}
	\caption{Branching ratio 
		$\mathcal{B}r (B_s \rightarrow \nu_C \nu_D)$ dependence on neutrino masses 
		$m_{\nu_C}$ and $m_{\nu_D}$ assuming $(V^\dag V)_{kk'}$ matrix element is real (left) and $(U^\dag V)_{jk}$ imaginary (right). See text for details.}
	\label{fig: Bnunu_CD}
\end{figure}

\begin{figure}[h]
	\includegraphics[scale=1.0]{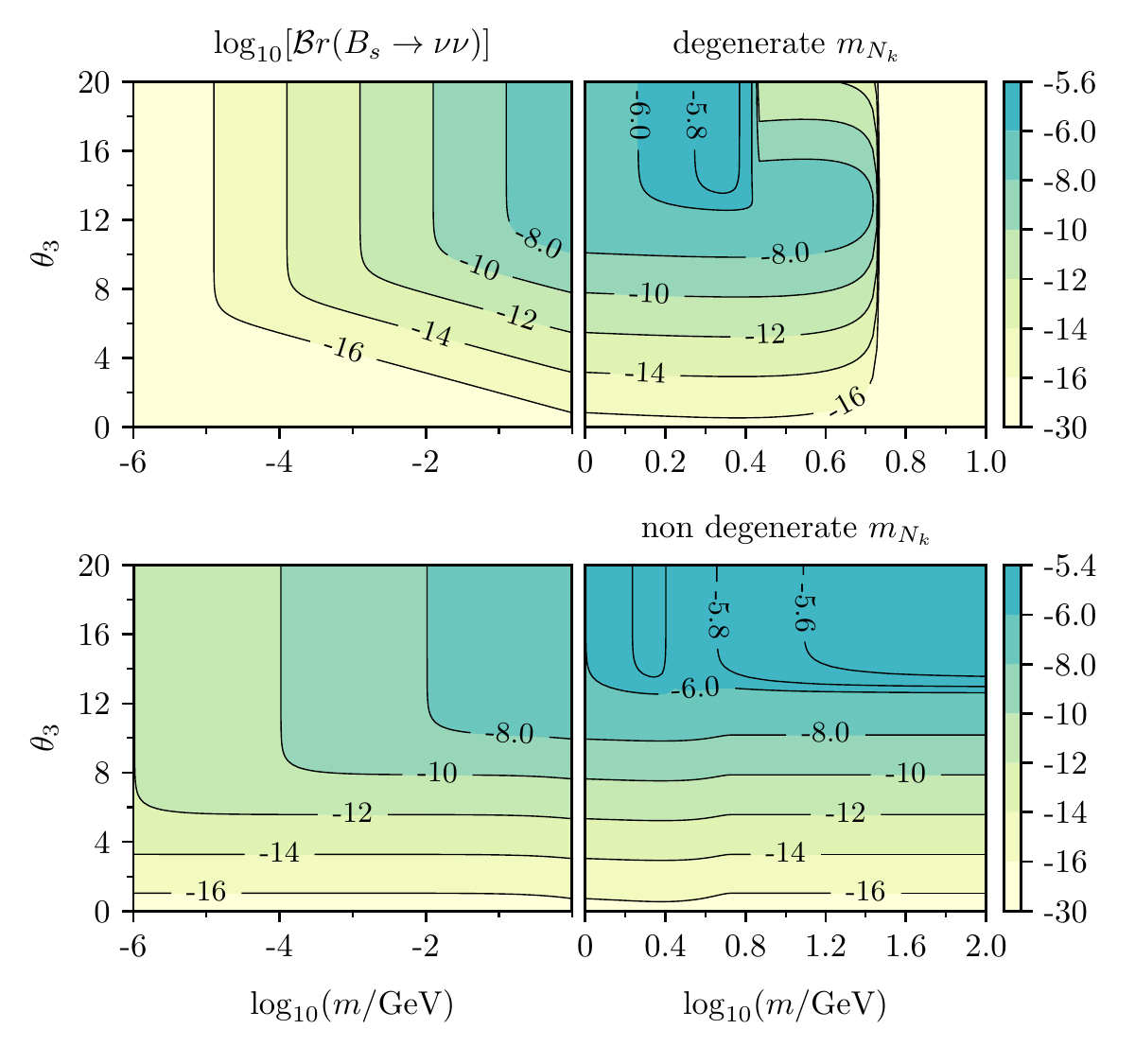}
	\caption{The $\ B_s \rightarrow \nu \nu$ branching ratio
		as a function of $\theta_3$ and $N^M_k$ neutrino mass $(m)$ in scenario with degenerate
		$N^M_k$ neutrino masses with $R = R(\theta_3)$ (upper plots), and in scenario with 
		$m_{N_1} = m_{N_2} = 2 \text{ GeV}$, $m = m_{N_3}$ and $R = R(\theta_3)$ (lower plots).
	    In both scenarios parameters $\theta_1$, $\theta_2$, $\phi_k$ for all $k$ are zero.
	}
	\label{fig: Bnunu}
\end{figure}

The $P \rightarrow \nu_C \nu_D$ decay is helicity suppressed and
therefore highly sensitive to neutrino masses as can be seen from Eq.~(\ref{eq: DW}).
Assume for the moment that $U$ and $R$ matrices are purely real or have a negligibly small imaginary part. 
Then the dependence of the decay width on $\mathcal U$ matrix elements can be factorized such that $\tilde {\mathcal B}r (P \rightarrow \nu_C \nu_D) \equiv  {\mathcal B}r (P \rightarrow \nu_C \nu_D)  /|\mathcal U_{CD}|^2$ becomes independent of $|\mathcal U_{CD}|$ and its dependence on neutrino masses $m_{\nu_C}$ and
$m_{\nu_D}$ is shown in Figure \ref{fig: Bnunu_CD}.
Note the different kinematical behavior of  $P\to N^M_k N^M_{k'}$ and $P\to \nu_j^M N_k^M$ due to the the sign difference  ${\rm sgn [Re}( (U^\dagger V)^2_{jk}) ] = - {\rm sgn [Re}((V^\dagger V)^2_{k'k''}) ]$, see Eq.~\eqref{eq: DW}\,.
Furthermore, assuming that neutrinos are (nearly) degenerate $D_N \approx m_N I$, $D_\nu \approx m_\nu I$ with $m_N \gg m_\nu$, we can decompose the branching fraction of
the $P \rightarrow \nu \nu$ decay into
\begin{equation}
\begin{split}
\mathcal{B}r(P \rightarrow \nu \nu) \approx& \
||U^\dag U||^2 \ \tilde{\mathcal{B}r}(P \rightarrow \nu^M_j \nu^M_{j'}) \ +\\[2mm] 
&+ 2 \  ||U^\dag V||^2 \ 
\tilde{\mathcal{B}r}(P \rightarrow \nu^M_j N^M_{k}) \ +\\[2mm] 
&+   ||V^\dag V||^2 \ 
\tilde{\mathcal{B}r}(P \rightarrow N^M_{k} N^M_{k'})\,.
\end{split}
\end{equation}
In this scenario, norms satisfy scaling relations described in subsection \ref{subsec scaling}.
From here it is easy to see that within these assumptions the branching ratio 
$\mathcal{B}r(P \rightarrow \nu \nu)$ is largest if $m_N$ is of a non-negligible fraction of the $P$ mass and the
$\xi$ scale is sufficiently large.  
This can be seen from Figure \ref{fig: Bnunu} for the case of $B_s$ decays. Due to scaling relations similar behavior like in Figure \ref{fig: Bnunu} is expected in case $\theta_3$ is exchanged by $\sum_k \theta_k$\,.

In general the branching fractions $\mathcal{B}r(P \rightarrow \nu \nu)$ are minimal 
in the Dirac limit. Precise values depend on the light neutrino masses, but are in any case negligibly small compared to experimental resolution of any currently foreseen experiments~\cite{Badin:2010uh}.
On the other hand, the maximal values of $\tilde{\mathcal{B}r}(P \rightarrow \nu_C \nu_{C'})$  are reached at $m_{N_k} = m_{N_k'} \approx 0.4\,m_P$ for $P \to N^M_k N^M_{k'}$, and at $m_{N_k} \approx 0.6\,m_P$ for $P \to \nu^M_j N^M_{k}$. In particular we find $\tilde{\mathcal{B}r}(B_s \rightarrow N^M_k N^M_{k'})_{\rm max} \simeq  3.5 \times 10^{-6}$, $\tilde{\mathcal{B}r}(B_d \rightarrow N^M_k N^M_{k'})_{\rm max} \simeq  1.1 \times 10^{-7}$ at $m_{N_k} =m_{N_{k'}} \simeq 2.2\,$GeV, and $\tilde{\mathcal{B}r}(B_s \rightarrow \nu^M_j N^M_{k})_{\rm max} \simeq  1.3 \times 10^{-6}$, $\tilde{\mathcal{B}r}(B_d \rightarrow \nu^M_j N^M_{k})_{\rm max} \simeq  4.2 \times 10^{-8}$ at $m_{N_k} \simeq 3\,$GeV\,.
The maximal values of $\mathcal{B}r(P \rightarrow \nu \nu)$, however, also crucially depend on the experimentally allowed values of $\mathcal U$ matrix elements. It turns out that currently the constraints are mildest for $m_{N_k}$ in the range of a few GeV~\cite{Drewes:2015iva}, in particular, there $|(U^\dag V)_{jk}|^2 < 5 \times 10^{-5}$ as reported by the DELPHI collaboration~\cite{DELPHICollaboration1997}\,.  This bound implies $\xi \ll 1$ and in turn $||V^\dag V|| \ll ||U^\dag V||$\,. If we thus assume approximately degenerate $N_k$ with masses $m_{N_k} \simeq 0.6 M_B \sim 3 \text{ GeV}$, 
and all $(U^\dag V)_{jk}$ matrix elements saturating the current experimental bound in that mass region we obtain 
\begin{subequations}
\begin{align}
\label{eq: Bs_nunu_max}
\mathcal{B}r(B_s \rightarrow \nu \nu) & \simeq  2 ||U^\dag V||^2 \tilde {\mathcal{B}}r  ( B_s \rightarrow \nu^M_j N^M_k)_{\rm max} <  1.2 \times 10^{-9}\,,  \\
\label{eq: Bd_nunu_max}
\mathcal{B}r(B_d \rightarrow \nu \nu) & <   3.8 \times 10^{-11}\,.
\end{align}
\end{subequations}

\subsection{$P \rightarrow \nu \nu \gamma $}
\label{sec:Pnunugamma}

\begin{figure}[h]
	\includegraphics[scale=1.0]{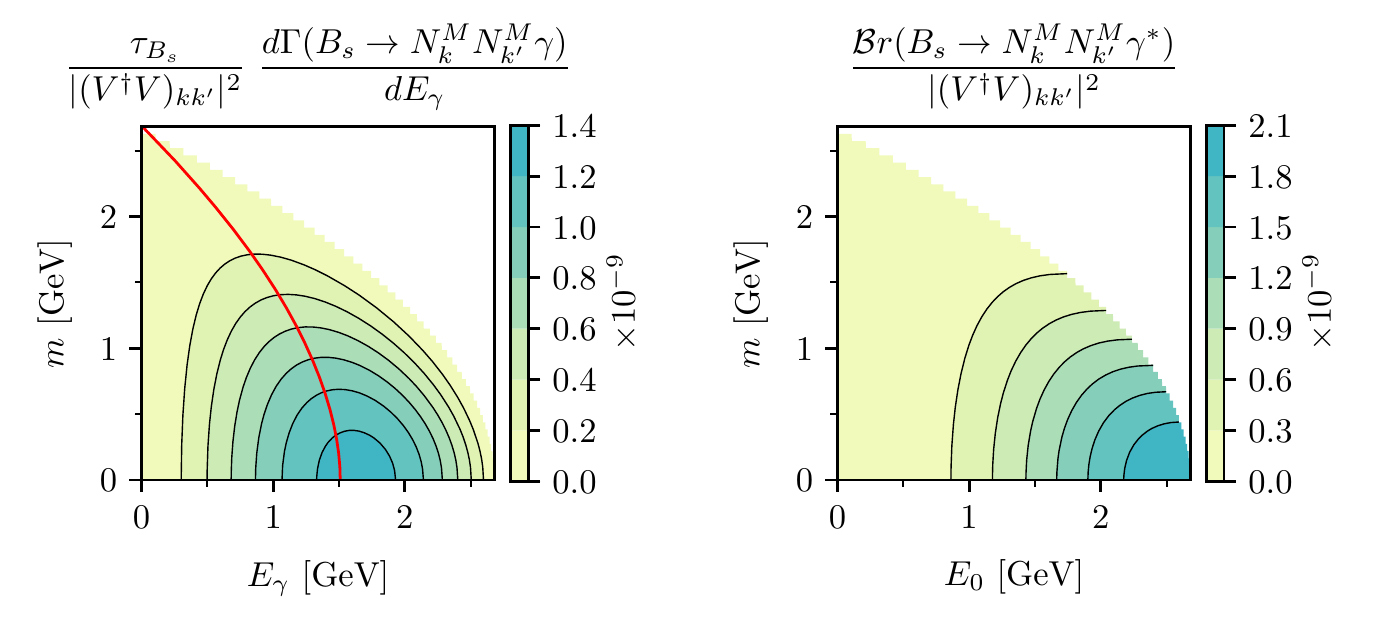}
	\caption{Spectrum of the $B_s \rightarrow N^M_k N^M_{k'} \gamma$
		decay as a function of the photon energy $E_\gamma$ and neutrino mass $m$ (left).
		The branching fraction of the $B_s \rightarrow N^M_k N^M_{k'} \gamma^*$ decay as a function of $m$ and photon threshold energy $E_0$ (right).
		In both plots $m = m_{N_k} = m_{N_{k'}}$ and matrix element 
		$(V^\dag V)_{kk'}$ is assumed to be real. 
		The red curve marks the average value of the photon energy $\langle E_\gamma \rangle$ in the decay
		as a function of $m$. 
	}
	\label{fig: Bnunugamma}
\end{figure}

In the limit $m_C, m_D \ll m_P$, the branching ratio $\mathcal{B}r( P \rightarrow \nu_C \nu_D \gamma)$ is approximately independent of neutrino masses
\begin{equation}
\mathcal{B}r( P \rightarrow \nu_C \nu_D \gamma)   \simeq 
\mathcal{B}r( P \rightarrow \nu_C \nu_D \gamma) \Big|_{m_C = m_D = 0}.
\end{equation}
 From here the identity  
$\mathcal{B}r( P \rightarrow \nu_C \nu_D \gamma) \simeq 
|\mathcal{U}_{CD}|^2 \ \tilde{\mathcal{B}}r( P \rightarrow \nu_C \nu_D \gamma)$,
 where $\tilde{\mathcal{B}}$ does not depend on $\mathcal{U}_{CD}$, holds.
Therefore in the case $|| D_N ||  \ll m_P$ the branching ratio 
of the $P  \rightarrow \nu \nu \gamma$ decay is given by
\begin{equation}
\begin{split}
\mathcal{B}r (P \rightarrow \nu \nu \gamma) =& 
\sum_{C,D} \mathcal{B}r( P \rightarrow \nu_C \nu_D \gamma )\\[1mm]
\simeq & \sum_{C,D} |\mathcal{U}_{CD}|^2 \
\tilde{\mathcal{B}}r( P \rightarrow \nu_C \nu_D \gamma )\Big|_{m_C = m_D = 0}\\[1mm]
\simeq& \ 3 \ \tilde{\mathcal{B}}r(P \rightarrow \nu_C \nu_D \gamma )
\Big|_{m_C = m_D = 0},
\end{split}
\end{equation}
which coincides with the SM prediction $\mathcal{B}r (P \rightarrow \nu \nu \gamma)^{\text{SM}}$ (with three massless neutrinos). In the last step the identity in~Eq.~(\ref{eq: SumUcd}) was used.
The $\tilde{\mathcal{B}}r(P \rightarrow \nu_C \nu_D \gamma )$ function takes lower
values if neutrino masses $m_{\nu_C}$ and $m_{\nu_D}$ increase as can be seen from the left-hand side plot in Figure 
\ref{fig: Bnunugamma} for the case of the $B_s \to N_k N_{k'} \gamma$ decay where we plot the photon energy spectrum of the decay (normalized to the $B_s$ lifetime and the relevant $|\mathcal U|^2$ matrix element) as a function of the neutrino mass. 

{
Using properties of the $||UU^\dag||$ norm described in Appedix~\ref{appendixB} one finds
that in cases $n_\nu = 3$ and $n_N \in \{2,3\}$ with $\max(D_\nu)/\min(D_N) \ll 1$ and
$\max(D_\nu) \ll m_P$, the branching ratio of 
$P \rightarrow \nu \nu \gamma$ decay lies in the interval:
}
\begin{equation}
 \label{eq: 2/3ineq}
\frac{2}{3} \ \leq \ \frac{ \mathcal{B}r (P \rightarrow \nu \nu \gamma)}{
\mathcal{B}r (P \rightarrow \nu \nu \gamma)^{\text{SM}}} \  \leq \ 1\,.
\end{equation}
The maximal value of the branching ratio takes place in the SM.
By taking $||D_\nu|| = ||D_N|| \rightarrow  0$, form factors $F_A(q^2)$ and $F_V(q^2)$
from the most recent estimate~\cite{Kozachuk:2017mdk} and integrating Eq.~(\ref{eq: partialG}) over
the whole phase-space, we find the SM predictions for branching ratios 
$\mathcal{B}r ( B_{s,d} \rightarrow \nu \nu \gamma)$ to be
\begin{subequations}
\begin{align}
\label{eq: Bs_nunugamma_SM}
\mathcal{B}r(B_s \rightarrow \nu \nu \gamma)_{\text{SM}}
&= 6.2 (1.9) \times 10^{-9}\,,\\[1mm]
\label{eq: Bd_nunugamma_SM}
\mathcal{B}r(B_d \rightarrow \nu \nu \gamma)_{\text{SM}}  
&= 2.8 (8) \times 10^{-10}\,,
\end{align}
\end{subequations}
where the $\mathcal O(30\%)$ uncertainties are dominated by the relevant hadronic form factor estimates, see also Appendix~\ref{Appendix}. 
As can be seen from Table~\ref{tab: Br_others}, the most recent form factor inputs lead to somewhat reduced predictions compared to previous estimates.
\begin{table}[h!]
	\centering
	\caption{SM predictions of  branching ratios of decays 
		$\{ B_s, B_d\} \rightarrow \nu \nu \gamma$ decays
		from other works. }
	\setlength{\tabcolsep}{12pt}
	\renewcommand{\arraystretch}{1.6}
	\begin{tabular}{  l  l  l r }
		\hline \hline
		$\mathcal{B}r(B_s \rightarrow \nu \nu \gamma)$  &
		$\mathcal{B}r(B_d \rightarrow \nu \nu \gamma)$ &
		year & reference\\   \hline 
		$6.2 \times 10^{-9}$          & $2.8 \times 10^{-10}$    & 2020 & this work  \\
		$3.68 \times 10^{-8}$     & $1.96 \times 10^{-9}$ & 2010 & \cite{Badin:2010uh}\\
		$1.2 \times 10^{-8}$       & not predicted                & 2002 & \cite{Cakir:2002xd} \\
		$1.8 \times 10^{-8}$       & $2.4 \times 10^{-9}$   & 1996 & \cite{LU1996348}   \\ 
		$7.5 \times 10^{-8}$       &  $4.2 \times 10^{-9}$  & 1996 & \cite{Aliev:1996sk}    \\[0.5mm] \hline \hline
	\end{tabular}
	\label{tab: Br_others}
\end{table}

In addition to the branching ratio, measuring the photon energy spectrum in $P\to \nu\nu\gamma$ decays would in principle allow to infer on the  mass spectrum of the neutrinos appearing in the final state. In particular, the average photon energy $\langle E_\gamma\rangle$, defined as
 \begin{equation}
\langle E_\gamma\rangle  = \frac{1}{\Gamma(P \rightarrow \nu \nu \gamma)}
    \ \int^{}_{} \frac{d \Gamma(P \rightarrow \nu \nu \gamma)}{d E_\gamma} E_\gamma d E_\gamma\,,
 \end{equation}
is inversely correlated with final state neutrino masses, as can be seen from the left-hand side plot in Fig.~\ref{fig: Bnunugamma}\,.

\begin{figure}[h!]
	\includegraphics[scale=1.0]{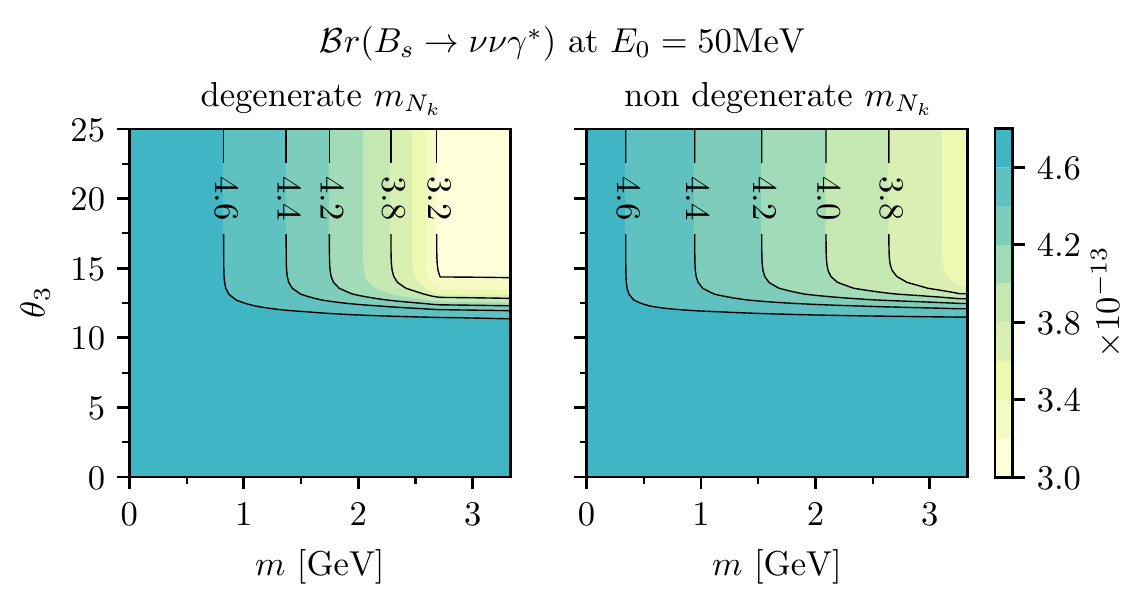}
	\caption{The branching fraction of the $B_s \rightarrow \nu \nu \gamma^*$ decay with a soft photon ($E_\gamma < E_0$) as a function of $\theta_3$ and  $(m)$ in scenario with degenerate $N^M_k$ neutrino masses with $R = R(\theta_3)$ (right plot), and in scenario with  $m_{N_1} = m_{N_2} = 2 \text{ GeV}$, $m = m_{N_3}$ and $R = R(\theta_3)$ (left plot).
	In both scenarios parameters $\theta_1$, $\theta_2$, $\phi_k$ for all $k$ are zero.	}
	\label{fig: Bnunu_gamma}
\end{figure}

Assuming neutrinos are completely unobserved, the $P \rightarrow \nu \nu \gamma$
decay contributes also to the invisible $P$ decay width effectively due to the finite resolution of any electromagnetic
 calorimeter, since photons with energies lower than some threshold energy $E_0$ of the detector are not registered. This contribution is simply given by
 \begin{equation}
\mathcal{B}r(P \rightarrow \nu \nu \gamma^*) =
   \tau_{P} \ \int^{E_0}_{0} \frac{d \Gamma(P \rightarrow \nu \nu \gamma)}{d E_\gamma} d E_\gamma\,,
 \end{equation}
where the other integrals are performed over the whole available phase space. On the right-hand side plot in Fig.~\ref{fig: Bnunugamma} we show the threshold energy and neutrino mass dependence of the $P \rightarrow \nu \nu \gamma^*$ decay branching fraction for the case $B_s \rightarrow N^M_k N^M_{k'} \gamma^*$. Specifically, in the SM and for a threshold energy of $E_0=50$\,MeV we obtain the predictions $\mathcal{B}r(B_s \rightarrow \nu \nu \gamma^*)^{E_0=50\,\rm MeV}_{\text{SM}} = 4.7 \times 10^{-13}$ and $\mathcal{B}r(B_d \rightarrow \nu \nu \gamma^*)^{E_0=50\,\rm MeV}_{\text{SM}} = 2.5 \times 10^{-14}$\,.  We show  in Fig.~\ref{fig: Bnunu_gamma} a typical dependence of the $B_s \rightarrow \nu \nu \gamma^*$ branching fraction on the neutrino mixing parameters in two scenarios for $N^M_k$ neutrino masses, $E_0 = 50 \text{ MeV}$ and with $R = R(\theta_3)$, $\theta_1 = \theta_2 = 0$. In comparison to Fig.~\ref{fig: Bnunu} we observe that these contributions to $B_s \to E_{\rm miss}$ are generically still more than three orders of magnitude smaller than the current upper limit on $B_s \rightarrow \nu \nu$ and thus completely subleading. However, at the same time, $P\to \nu\nu \gamma^*$  are always much bigger than $P \to \nu\nu\nu\nu$~\cite{Bhattacharya:2018msv} and are thus expected to dominate $P\to E_{\rm miss}$ in the (pseudo)Dirac neutrino limit. In addition,  $P \rightarrow \nu \nu \gamma$ might contribute effectively to $P\to E_{\rm miss}$ also due to other detector effects, such as non-perfect $4\pi$ coverage. Such contributions are however difficult to model without a detailed knowledge of the detector components and geometry and we leave such a study to the experimental collaborations. 

\section{Conclusions}\label{conclusions}


In this work we have reconsidered Majorana neutrino mass models and derived a model-independent general 
parametrization of neutrino mass matrices with physically interpretable and irreducible set of parameters.
The parametrization is valid for any number of left-handed (as in SM) and right-handed (gauge singlet Majorana) neutrinos and
for all mass hierarchies.
In particular, in the heavy Majorana neutrino limit we recover the standard Casas-Ibara parametrization~\cite{Casas:2001sr},
while the parametrization nicely interpolates also through the (pseudo)Dirac neutrino limit.

We have applied the new parametrization to the study of  $P \rightarrow \nu \nu$ and $P \rightarrow \nu \nu \gamma$  decays within the SM extended by $n_N = 3$ additional 
singlet neutrinos.\footnote{{The parametrization can trivially be applied to other neutral current mediated processes involving pairs of neutrinos. In the case of charged current mediated processes commonly used to constrain the PMNS matrix (see e.g. Ref.~\cite{Fernandez-Martinez:2016lgt}), or in searches for massive Majorana neutrinos with lepton number violating signatures~\cite{Atre:2009rg, Abada:2017jjx}, one instead needs to consider additional dependence on the unitary $O_L$ matrix.}}
Along the way we have updated the SM predictions for the branching ratios of $B_{s,d} \rightarrow \nu \nu \gamma$ decays and found almost an order of magnitude smaller values compared to previous estimates, mainly due to a recent reevaluation of the relevant hadronic form factors. 

Finally, we have discussed the sensitivity of the $B_{s,d} \rightarrow E_{\rm miss} (\gamma)$ decays
to neutrino mass and mixing parameters. In the case of $B_{s,d} \rightarrow E_{\rm miss}$, for typical EM calorimeter threshold energies and assuming $4\pi$ coverage, the dominant contribution could still come from $B_{s,d} \rightarrow \nu \nu$ decays where one of the final state neutrinos is predominantly a SM gauge singlet of mass of the order a few GeV. However, the maximum allowed branching ratios, given by the current experimental bounds on the relevant neutrino mixing matrices, are at least four orders of magnitude below the direct limits from the B factories. It remains to be seen if Belle II can reach the required sensitivity to constrain the parameter space of neutrino mass models in this interesting region. 

 In the case of $B_{s,d} \rightarrow E_{\rm miss} \gamma$ decays, additional light neutrinos in the final state could affect both the branching ratios as well as the photon energy spectra and thus in principle allow to extract information on the neutrino mass parameters.
 Unfortunately, however,  possible deviations from SM predictions (i.e. the limit of massless neutrinos) are theoretically constrained and  at most comparable to current uncertainties due to the limited knowledge of the relevant form factors. Any relevant experimental sensitivity to neutrino mass parameters is thus conditional upon an improved understanding of the relevant hadronic parameters, which could possibly come from future Lattice QCD studies (see Refs.~\cite{Kane:2019jtj, Desiderio:2020oej} for current prospects). 

\begin{acknowledgments}
{
We are grateful to Klemen \v Sivic for his help in the derivation of the  lower bound on the $|| UU^\dag ||$ norm,
and to Miha Nemev\v sek for his comments on the manuscript. JFK acknowledges the financial support
from the Slovenian Research Agency (research core funding No. P1-0035 and J1-8137).}
\end{acknowledgments}

\bibliographystyle{elsarticle-num}
\bibliography{ref_param}

\appendix
\section{Calculation of $P \rightarrow \nu \nu$ and $P \rightarrow \nu \nu \gamma$ decay widths }
\label{Appendix} 

Below we summarize our calculation of the $B_{s,d} \rightarrow \nu \nu (\gamma)$ decay widths as discussed in the main text.
We consider the physically relevant scenario with $n_{\nu} = 3$ neutrinos $\nu^M_j$ and
$n_N$ neutrinos $N^M_k$. The Majorana fields are defined in a way such that
the Dirac limit can be approached analytically with both $D_\nu$
and $D_N$ matrices being positive semi-definite. For other choices of the relecant phase factors
one should properly redefine the mixing matrix $\mathcal{U}$. Note that our calculation can be applied also to the corresponding $K, D$ meson decays, with suitable quark flavor replacements.

We calculate the decay widths $B_{s,d} \rightarrow \nu \nu(\gamma)$ using the relevant effective weak Lagrangian~\cite{Inami:1980fz, Buchalla:1993bv}
\vspace{2mm}
\begin{equation}\label{Leff}
\mathcal{L}_{\rm{eff}} =  
\frac{4 G_F \alpha_{\text{em}}}{2\sqrt{2} \pi \sin^2{\theta_W}} \sum_{q=s,d}
V_{tq}^* V_{tb} X(x_t) \left[ \overline{b} \gamma_\mu P_L q\right]
\sum^{3+n}_{C,D=1} J^{\mu}_{CD}\,,
\end{equation}
where the leptonic current $J^{\mu}_{CD}$ is given by
\begin{subequations}
\begin{align}
\label{Jcd}
J^\mu_{CD}& =
\bra{\nu_C(\vec{p}_C), \nu_D(\vec{p}_D)}
\sum_{a=1}^3 \overline{\nu_{aL}} \gamma^\mu  \nu_{aL} \ket{0} \thinspace 
\text{e}^{-\text{i}(p_C+p_D)x},
\\[2mm]
\label{Jcd2}
& = - \ \mathcal{U}_{CD}\thinspace  [\overline{u}_C \gamma^\mu P_L v_D]
+ \thinspace \thinspace \mathcal{U}^*_{CD} \thinspace 
[\overline{u}_D \gamma^\mu P_L v_C]\,.
\end{align}
\end{subequations}
The relevant loop function $X(x_t)$ can be written as
\begin{equation}
\label{Xfull}
X(x_t, x_\mu) =   X_0(x_t) + \frac{\alpha_s(\mu)}{4 \pi} X_1(x_t,x_\mu)\,,
\end{equation}
where $X_0(x_t)$ is the Inami-Lim function~\cite{Inami:1980fz}
\begin{equation}
\label{X0}
X_0(x_t) = \frac{x_t}{8} \left[  \frac{x_t+2}{x_t-1}+\frac{3(x_t-2)}{(x_t-1)^2} \ln{x_t} \right],
\end{equation}
and the leading QCD corrections are parametrized by $X_1$ whose explicit expression can be found in Refs.~\cite{Buchalla:1993bv, Buras:1998raa}. 
Here $x_\mu = \mu^2/M^2_W$,  $x_t = m_t^2/M^2_W$ and the $\overline{\text{MS}}$ QCD renormalization scheme is assumed throughout. In the following we compress the common constant prefactors entering the Lagrangian into
\begin{equation}
C\equiv \frac{G_F \alpha_{\mathrm{em}} }{2\sqrt{2}\pi \sin^2{\theta_W}} 
V_{tq}^* V_{tb} X(x_t)\,.
\end{equation}
Above and in the following we have suppressed the light flavor ($q=s,d$) indices where the identification of the relevant $B_{(q)}$ meson flavor is unambiguous.

For the $P \rightarrow \nu \nu$ decay we parametrize the relevant hadronic matrix elements in the standard way
\begin{subequations}
\begin{align}
\bra{0}\bar{b} \gamma^{\mu} q\ket{P(p)}&=0\,,\\[1mm]
\bra{0}\bar{b} \gamma^{\mu} \gamma^5 q\ket{P(p)}&=\mathrm{i} f_{P} \thinspace p^{\mu}\,,
\end{align}
\end{subequations}
where $f_P$ is the relevant $P=B_{s,d}$ meson decay constant. In particular,
we use $f_{B_s} =  {224}\rm{MeV}$ and $f_{B_d} =  {186}\rm{MeV}$  from Ref.\cite{Dowdall:2013tga}.  

In the case of the radiative decay,
only the emission of photons from the hadronic part is relevant and is parametrized by the relevant radiative form factors
\begin{subequations}
\begin{align}\label{vak1}
\langle \gamma(k) | \bar{b} \gamma_\mu q | P (k+q) \rangle &= e \epsilon_{\mu \nu \rho \sigma}
\epsilon^{* \nu} q^\rho k^\sigma \; \frac{F_V^{}(q^2)}{m_{P}}\,, \\
\label{vak2}
\langle \gamma(k) | \bar{b} \gamma_\mu \gamma_5 q | P (k+q) \rangle &=-\mathrm{i} e  \left[ \epsilon_\mu^* (k q)
-(\epsilon^* q) k_\mu \right] \frac{F_A^{}(q^2)}{m_{P}}\,.
\end{align}
\end{subequations}
Here $q = p_C + p_D = p - k$ and  $q^2 =m_P^2 - 2 m_P E_\gamma$.
For the axial (A) and vectorial (V) form factors $F_A^{}(q^2)$ and $F_V^{}(q^2)$ we take the most recent estimate~\cite{Kozachuk:2017mdk} parametrized by
\begin{equation}\label{form}
F^{}_X (q^2) = \frac{F (0)}{ (1 - q^2/M^2_R) 
	\left[1 - \sigma_1 \thinspace (q^2/M^2_R)+ \sigma_2 \thinspace (q^2/M^2_R)^2 \right]}\,,
\end{equation}\\
where $F(0)$, $\sigma_1$, $\sigma_2$ and $M_R$ parameters for $P=B_{s,d}$ are given in Ref.~\cite{Kozachuk:2017mdk}.
After a quick calcuation one finds the expression for the
$P \rightarrow \nu_C \nu_D$ decay width
\begin{equation}\label{eq: DW}
\begin{split}
\Gamma(P \rightarrow \nu_C \nu_D) =
16 |C|^2 f^2_{P} \Bigg\{&
\frac{1}{2} | \mathcal{U}_{CD}|^2 
\left[ m^2_P \thinspace (m^2_C + m^2_D)  -  (m^2_C - m^2_D)^2  \right]+  \\
& +  \thinspace m_C m_D m^2_P \thinspace 
\text{Re} \big( \mathcal{U}^2_{CD} \big) 
\Bigg\}  \frac{Q}{2m_P}\,,
\end{split}
\end{equation}
where
\begin{equation}\label{Q2final}
Q = 
\frac{ 1}{(4 \pi) m_P} \sqrt{ \left( \frac{m^2_P +m^2_C -m^2_D}{2m_P} \right)^2 -m^2_C}
\times \left\{
\begin{array}{r l l}
1 &: &\mathrm{Dirac},\\[1mm]
\frac{1}{2} &: &\mathrm{Majorana}\,.
\end{array} \right. 
\end{equation}
In the Dirac limit the process $P \rightarrow \nu^M_j N^M_j$ is forbidden.
Similarly, the triply differential $P \rightarrow \nu_C \nu_D \gamma$ decay width (for Majorana neutrinos) is given by
\begin{equation}
\label{eq: partialG}
\begin{split}
\frac{d^3 \Gamma(P \rightarrow \nu_C \nu_D \gamma)}
{d E_\gamma d E_C d \Omega}
=& \frac{ |C|^2   \alpha_{\mathrm{em}}}{ 4 \pi^3 m_P}
\left[  {F_A}^2 + {F_V}^2 \right] 
\Bigg\{ -   \thinspace m_C m_D E^2_\gamma \thinspace
\text{Re} \big( \mathcal{U}^2_{CD} \big) +\\[2mm]
&+
|\mathcal{U}_{CD}|^2  \bigg[
(k \cdot p_C)^2  - E_\gamma (2 E_C +E_\gamma)  (k \cdot p_C) + E^2_\gamma  E_C m_P
\bigg] 
\Bigg\} \times\\[2mm]
&\enspace \enspace \times
\theta(m_{P}-E_C-E_\gamma) \thinspace \theta(E_C-E_{C\rm{min}}) \thinspace
\theta(E_{C\rm{max}}-E_C)\,.
\end{split}
\end{equation}
Note that due to Majorana nature of neutrinos the full integral over the solid angle $d \Omega$ gives $2\pi$
instead of the usual $4\pi$. From kinematic constraints one furthermore obtains
\begin{equation}\label{Ec}
{E_C}_{\rm{min}}^{\rm{max}}=\frac{m_{P}-E_\gamma}{2} \Delta  \   \pm \  
 \frac{1}{2} E_\gamma 
\sqrt{\Delta^2-\frac{4m^2_{C}}{m^2_{P} - 2 m_{P} E_\gamma} }\,,
\end{equation}
where is $\Delta$ equal to 
\begin{equation}
\Delta = 1+ \frac{m^2_C- m^2_D}{m^2_{P} - 2 m_{P} E_\gamma}\,,
\end{equation}
while $k \cdot p_C$ is given by
\begin{equation}
k \cdot p_C =\frac{1}{2} \left( -m_P^2 -m^2_C + m^2_D + 2 m_P E_\gamma + 2 m_P E_c \right)\,.
\end{equation}

Finally, in our numerical results we use $\mu = m_Z = {91.2}\rm{GeV}$, 
$\alpha_{\text{em}} = 1/137$, 
$m_{B_s} = {5.37}\rm{GeV}$, 
$m_{B_d} = {5.28}\rm{GeV}$,
$m_W = {80.4}\rm{GeV}$,
$|V_{tb} V_{ts}^*| = 0.0403$, 
$|V_{tb} V_{td}^*| = 0.00875$, $\sin^2{(\theta_W)}=0.22$, 
$\tau_{B_s} =1.51 \text{ ps} $, $\tau_{B_d} = 1.52 \text{ ps} $, $\alpha_s (m_Z) = 0.118$ and
$m_t(m_Z) = {172}\rm{GeV}$~\cite{Tanabashi:2018oca}.

\section{Lower bounds on $||UU^\dag||$}
\label{appendixB}

In this section we formally prove Eq. (\ref{eq: 2/3ineq}) and discuss additional properties of the $||UU^\dag||$ norm. We assume a model with $n_\nu$ light neutrinos ($m_{\nu_j}/m_P \ll 1$) and $n_N$ heavy neutrinos ($m_{\nu_j}/m_{N_k} \ll 1$).  From properties of $P \rightarrow \nu \nu \gamma$ decay follows that inequality:
\begin{equation}
||UU^\dag||^2 \ \tilde{\mathcal{B}r}(P \rightarrow \nu_j \nu_j \gamma) \bigg|_{m_{\nu_j} = 0}
\leq  \ \mathcal{B}r(P \rightarrow \nu \nu \gamma),
\end{equation}
\noindent
holds. Moreover from upper limit of the branching ratio $\mathcal{B}r(P \rightarrow \nu \nu \gamma)$ we find:
\begin{equation}
\tilde{\mathcal{B}r}(P \rightarrow \nu_j \nu_j \gamma) \bigg|_{m_{\nu_j} = 0} = \frac{1}{n_\nu}
\mathcal{B}r(P \rightarrow \nu \nu \gamma)^{\text{SM}}.
\end{equation}
\noindent
This is a direct consequence of Eq. (32). From here, within assumed model we have:
\begin{equation}
\frac{||UU^\dag||^2 }{ n_\nu} \leq \frac{ \mathcal{B}r(P \rightarrow \nu \nu \gamma) }{ \mathcal{B}r(P \rightarrow \nu \nu \gamma)^{\text{SM}} } \leq 1. 
\end{equation}
\noindent
Value of the norm is due to derived parametrization of neutrino matrices directly related
to the eigenvalues $\mu_j$ of the $QQ^\dag$ matrix through relation:
\begin{equation}
\label{eq: UUlambda}
|| U^\dag U ||^2 = \sum^{n_\nu}_{j = 1} \frac{1}{(1 + \mu_j)^2}.
\end{equation}
\noindent
This equation is obtained by diagonalizing $QQ^\dag$ matrix inside Frobenius norm. For
$UV^\dag$ and $VV^\dag$ matrices similar formulas can be found:
\begin{align}
|| U^\dag V ||^2 = || 	V^\dag U ||^2 =& \sum^{n_\nu}_{j = 1} 
\frac{\mu_j}{(1 + \mu_j)^2}, \\[1mm]
|| V^\dag V ||^2  =& \sum^{n_\nu}_{j = 1} \frac{\mu^2_j}{(1 + \mu_j)^2}.
\end{align}
\noindent
We use Eq. (\ref{eq: UUlambda}) as a starting point to determine theoretical lower bounds on the $|| UU^\dag ||$ norm for different choices of $n_\nu$ and $n_N$.

\subsection{Case $n_{\nu} \geq n_N$}

In scenario with $n_{\nu} \geq n_N$ we use following theorems.
\begin{notation}
	Let $X$ be a $n \times n$ hermitian matrix, then we denote its eigenvalues as $\lambda_j(X)$, where $\lambda_1(X) < ... < \lambda_n(X)$. 
\end{notation}
\begin{definition} (Loewner order)
	Let $A$ and $B$ be hermitian matrices, then $A \leq B$ if and only if $A-B$ is positive semidefinite matrix.
\end{definition}
\begin{theorem} 
	\label{theorem 1}
	(Loewner order is compatible with congruence)
	If $A$ and $B$ are hermitian matrices and $A \leq B$, then for any matrix $X$: $XAX^\dag \leq XBX^\dag$.
\end{theorem}
\begin{theorem} 
	\label{theorem 2}
	(Wely’s monotonicity theorem, [\cite{10.5555/2422911}, Corollary 4.3.3]) 
	If $A$ and $B$ are $n \times n$ hermitian matrices with $A \leq B$, then 
	$\lambda_k(A) \leq  \lambda_k(B)$ for each $k = 1,..,n$.
\end{theorem}
\begin{theorem} 
	\label{theorem 3}
	([\cite{10.5555/2422911}, Theorem 1.3.20]) 
	Let $A$ be a $m \times n$ matrix and $B$ be a $n \times m$
	matrix with $m \leq n$. Then $BA$ matrix has the same eigenvalues as $BA$, 
	together with additional $n - m$ eigenvalues equal to 0.
\end{theorem}
We denote by $a$ the largest eigenvalue of $D_\nu$ and by $b$ the smallest eigenvalue of $D_N$,
therefore $ D^{-1}_N  \leq 1/b I$. Using Theorem (\ref{theorem 1}) we get:
\begin{equation}
QQ^\dag = (D^{1/2}_\nu P R) D^{-1}_N (D^{1/2}_\nu P R )^\dag \ \leq \ 
\frac{1}{b} D^{1/2}_{\nu} P RR^\dag P^\dag D^{1/2}_\nu.
\end{equation}
\noindent
By Theorem (\ref{theorem 3}) we find $\lambda_j(QQ^\dag) = 0$ for $j = 1,..., n_\nu - n_N$.
Next for $j > n_\nu - n_N$, we apply theorems (\ref{theorem 2}), (\ref{theorem 3}) and (\ref{theorem 1}) in this order to obtain:
\begin{equation}
\lambda_j(QQ^\dag) \ \leq  \ \frac{1}{b} \lambda_j((D^{1/2}_\nu P R)(R^\dag P^\dag D^{1/2}_\nu))
\ \leq \ \frac{1}{b} \lambda_j(R^\dag P^\dag D_\nu P R) \ \leq \ \frac{a}{b} \lambda_j(R^\dag R). 
\end{equation}
\noindent
From here, we finally get:
\begin{equation}
\label{eq: UUlambda1}
|| U^\dag U ||^2 \geq n_\nu - n_N +  \sum^{n_N}_{k = 1} \frac{1}{[1 + \frac{a}{b}\lambda_k(RR^\dag)]^2}.
\end{equation}
\noindent
From the property $(R^\dag R)^{-1} = (R^\dag R)^*$ follows directly, that if $\lambda(RR^\dag)$ is
an eigenvalue of the $R^\dag R$ matrix, then $1/\lambda(R^\dag R)$ is also eigenvalue of $R^\dag R$ matrix. A consequence of this property is that for any $(2n+1) \times (2n+1)$ orthogonal matrix $R$, the $RR^\dag$ matrix has at least one eigenvalue equal to 1. 
Therefore, if $R$ is a general $n \times n$ complex orthogonal matrix, then $R^\dag R$ matrix have maximally  $\lfloor n/2 \rfloor$ (integer part of $n/2$) eigenvalues which are greater 
than 1. This implies:
\begin{equation}
|| U^\dag U ||^2 \geq n_\nu - \lfloor n_N/2 \rfloor.
\end{equation}
\noindent
Using property:
\begin{equation}
\frac{1}{(1 + \frac{a}{b} \lambda)^2} + \frac{1}{(1 + \frac{a}{b} \frac{1}{\lambda})^2} \geq
\frac{1}{(1 + \frac{a}{b})^2},
\end{equation}
\noindent
a better lower bound can be obtained:
\begin{equation}
|| U^\dag U ||^2 \geq n_\nu - n_N +  \frac{ \lfloor \frac{n_N+1}{2} \rfloor }{(1 + \frac{a}{b})^2}.
\end{equation}
\noindent
Eigenvalues of $RR^\dag$ matrix depend only on $\theta_k$ parameters if $R$ matrix
is parametrized in the following way:
\begin{equation}
R( \underline{\phi}, \underline{\theta} ) = \prod^{n}_{k = 1} R_k(\phi_k)  \prod^{n}_{k = 1} R_{n+k} (\theta_k),
\end{equation} 
\noindent
where $R_1,..., R_n$ are real orthogonal matrices and therefore unitary. We can absorb them in the process of diagonalizing $RR^\dag$ matrix:
\begin{align}
\det \left( 
\prod^{n}_{k = 1} R_k(\phi_k)  
\prod^{n}_{k = 1} R_{n+k}(\theta_k) 
R^\dag_{n+k} (\theta_k) 
\prod^{n}_{k = 1} R^\dag_k(\phi_k)
- \lambda I \right)
= \det \left( \prod^{n}_{k = 1} R_{n+k}(\theta_k) 
R^\dag_{n+k} (\theta_k)  - \lambda I \right).
\end{align} 
\noindent
From here, eigenvalues of $RR^\dag$ matrix must depend only on $\theta_k$ parameters.

\subsection{Case $n_\nu < n_N$}

In scenario with $n_\nu < n_N$ is more difficult to obtain eigenvalues of $QQ^\dag$ matrix,
since $S$ matrix is present in it. In case $n_N \geq 2 n_\nu$ the lowest bound on $||UU^\dag||$ norm is 0. This can be proven using $R = I_{n_\nu \times n_\nu}$, $D_\nu = a I_{n_\nu \times n_\nu}$, $S = [I_{n_\nu \times n_\nu}, \mathrm{i} x I_{n_\nu \times n_\nu}, 
0_{(n_N - 2n_\nu) \times (n_N - 2n_\nu)}]$, where $x \in \mathcal{R}$ and:
\begin{equation}
D_N = \begin{pmatrix}
b I_{n_\nu \times n_\nu} & 0 & 0 \\
0 & c I_{n_\nu \times n_\nu} & 0 \\
0 & 0 & C_{(n_N - 2n_\nu) \times (n_N - 2n_\nu)} \\
\end{pmatrix}.
\end{equation}
\noindent
where $C$ is a positive semi-definite diagonal matrix. From here one gets:
\begin{equation}
QQ^\dag = a \frac{1 + x^2}{|b - c x^2|} I_{n_\nu \times n_\nu}.
\end{equation} 
\noindent
For $x \neq \pm \sqrt{b/c}$ matrix $SD_N S^T$ is invertible. When $x$ approaches to $\sqrt{b/c}$
eigenvalues of $QQ^\dag$ matrix become very large and therefore the lowest value for the $||UU^\dag||$ norm is 0.

\end{document}